\pdfoutput=1
\documentclass{JINST}
\usepackage{textcomp} 

\DeclareGraphicsExtensions{.pdf,.png,.jpg} 

\title{Design and construction of a Cherenkov imager for charge measurement of nuclear cosmic rays}
\author{O.~Bourrion$^a$\thanks{Corresponding author.}, Ch.~Bernard$^a$, 
D.~Bondoux$^a$, J.L.~Bouly$^a$, J.~Bouvier$^a$, B.~Boyer$^a$, M.~Brinet$^a$, M.~Buenerd$^a$, G.~Damieux$^a$, L.~Derome$^a$, L.~Eraud$^a$, R.~Foglio$^a$, D.~Fombaron$^a$, 
D.~Grondin$^a$, M.~H.~Lee$^d$, L.~Lutz$^d$, M.~Marton$^a$, A.~Menchaca-Rocha$^c$, A.~Pelissier$^a$, J.N.~P\'eri\'e$^b$, A.~Putze$^{a,1}$, S.~Roudier$^a$, Y.~Sallaz-Damaz$^a$, E.~S.~Seo$^d$, J.P.~Scordilis$^a$, Y.~S.~Yoon$^d$.\\
\llap{$^a$}Laboratoire de Physique Subatomique et de Cosmologie,\\ 
Universit\'e Joseph Fourier Grenoble 1,\\
  CNRS/IN2P3, Institut Polytechnique de Grenoble,\\
  53, rue des Martyrs, Grenoble, France\\
\llap{$^b$}Université de Toulouse; INSA, UPS, Mines Albi, ISAE; \\
  ICA (Institut Clément Ader); \\
 133, avenue de Rangueil, F-31077 Toulouse, France \\
\llap{$^c$} Instituto de Física, UNAM, A.P. 20-364, 01000 Mexico DF, Mexico\\
\llap{$^d$} U. Maryland, College Park MD 20742, USA\\
\llap{$^1$} Now at: The Oskar Klein Centre for Cosmoparticle Physics,\\
 Department of Physics, Stockholm University, AlbaNova,\\
 SE-10691 Stockholm, Sweden\\
 } 

\abstract{A proximity focusing Cherenkov imager called CHERCAM, has been built for the charge measurement of nuclear cosmic rays with the CREAM 
instrument. It consists of a silica aerogel radiator plane
across from a detector plane equipped with 1,600 1`` diameter photomultipliers. The two planes are separated by a
ring expansion gap. The Cherenkov light yield is proportional to the charge squared of the incident particle.
The expected relative light collection accuracy is in the few percents range. It leads to an expected single element
separation over the range of nuclear charge Z of main interest $1 \leq Z \lesssim 26$. CHERCAM is designed to fly with the CREAM balloon experiment. The design of the instrument and the implemented technical solutions allowing its safe operation in high altitude conditions (radiations, low pressure, cold) are presented.} 


\begin{document}


\section{Introduction}
The CREAM experiment investigates the nature and the origin of nuclear cosmic
rays (CR) by measuring the high energy CR flux at the statistical limit
accessible to the current generation of balloon experiments. The measurement of
the CR spectra of nuclear elements from proton to iron between $10^{10}$ eV
for Z$>$2 and $10^{12}$\,eV for H and He particles, up to $10^{15}$ eV, 
provides new data on their characteristics and abundances, including
measurement of the B/C secondary-to-primary ratio in this energy range. The
study of individual elemental fluxes allows to probe the current models
of acceleration mechanisms, and will provide clues for the interpretation of
the ``knee'' in the inclusive spectrum and for the physics of the Galactic CR 
transport.
The CREAM experiment includes a set of sub-detectors able to measure the particle
energy and charge (a more detailed description of the experiment can be found in \cite{Ahn}). 
The particle energy is measured by means of a
hadronic calorimeter with a nearly constant resolution over the three orders of
magnitude covered. The calorimeter can be combined and cross-calibrated with a
transition radiation detector (TRD) in some flight configurations. 
The cosmic-ray charge measurement is achieved by combining scintillation hodoscope, a Cherenkov threshold detector,
silicon micro-strip counters, and the Cherenkov imager optimized for charge
measurement, described in this paper (CHERCAM, or CHERenkov CAMera).
The purpose of the designed detector is the measurement of nuclear cosmic-ray
charges ranging from proton (Z=1) up to iron (Z=26). 
The interest of using the Cherenkov yield is that, it can provide a particle charge measurement with a
constant resolution $\Delta$Z through the whole range of charges 
\cite{AMScounterBu,AMScounterBarao,theseYO}, if systematic errors can be
properly controlled. 
The proximity focusing principle used for the design was chosen because of both its suitability to
the measurement purposes and the simplicity of the geometrical configuration. Additionally 
it can conveniently fulfill the constraints of an embarked experiment.
The photon detector plane is composed of 1,600 1`` diameter Photonis XP3112 photomultiplier tubes (PMT) arranged in a square pattern.
The ring expansion gap was chosen in order to allow a development of a Cherenkov ring with a
diameter larger than one single PMT, but only covering a small number of neighboring PMTs.
The simulation showed that such a configuration could provide a good
enough resolution to achieve the required single charge separation of elements of about 3 $\sigma$ RMS.

\section{Detector description}
\begin{figure}[th]
\begin{center}
\includegraphics[angle=-90,width=0.7\textwidth]{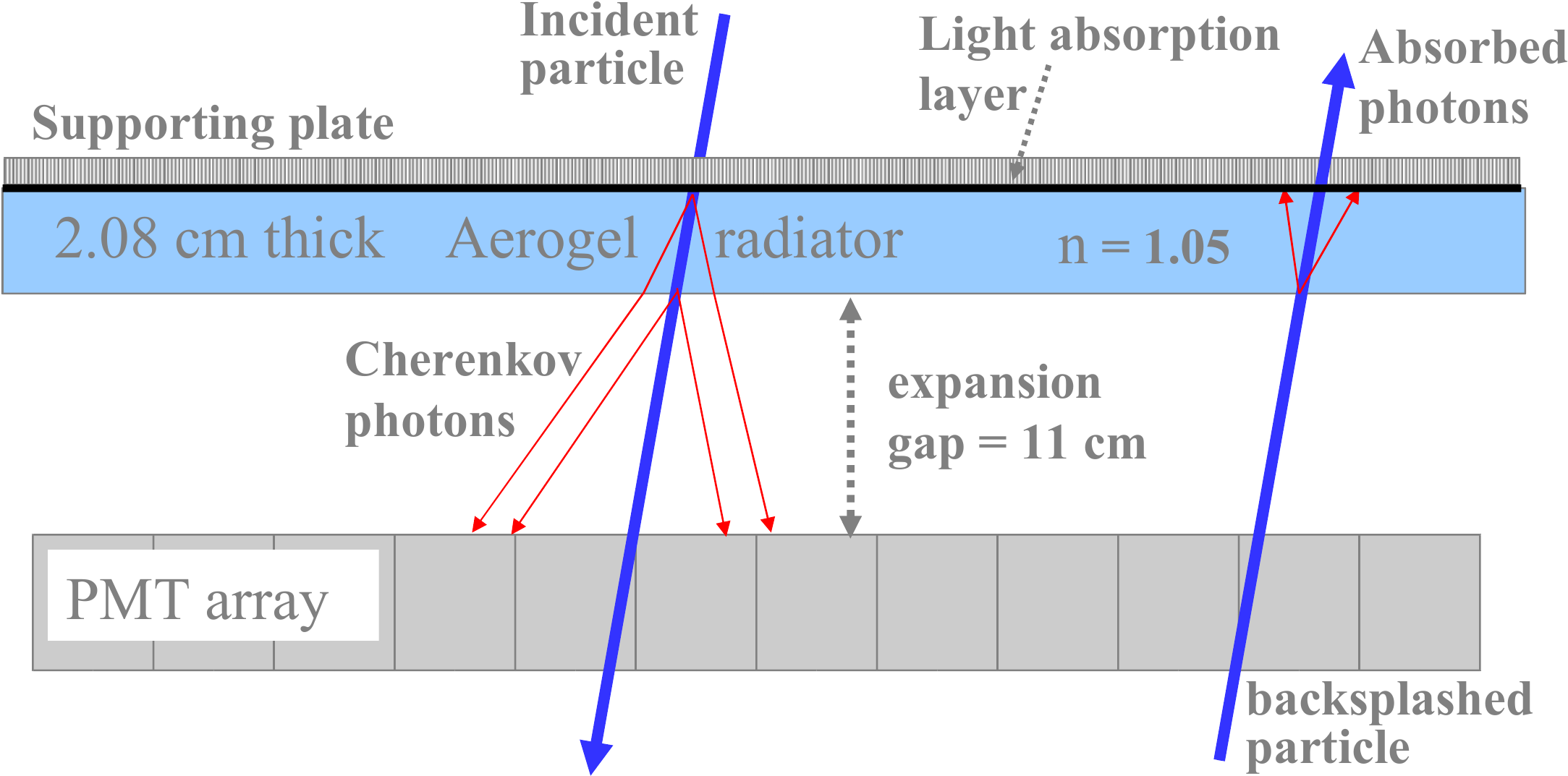}
\caption{Schematic view of the Cherenkov imager principle.}
\label{RichPrinciple}
\end{center}
\end{figure}

The counter architecture is derived from the solution developed for the AMS
imager \cite{AMScounterBu,AMScounterBarao}. The principle is illustrated in
figure \ref{RichPrinciple}. 
The radiator consists of silica aerogel plane equipped with  $\rm 10.5
\times 10.5\ cm^2$, 2.08\,cm thick tiles, with a refractive index n=1.05. The Cherenkov ring expansion
gap between radiator and detector plane is 11\,cm.
Many choices in CHERCAM's design were driven by cost considerations and constraint related to embarked
experiments (weight and power budget, simple and robust
technology, time budget).  The detector plane is an array of 1,600 1``
diameter Photonis XP3112 PMTs in a square arrangement
(i.e., not closely packed) with a 27.5\,mm pitch. This geometrical arrangement provides a 50\% active area. 
Therefore, the light collection suffers a significant dependence on the particle impact coordinates.
The impact of this effect on the charge resolution can be easily mitigated provided the hit coordinates are precisely known. 
An off-line likelihood analysis provides the most probable coordinate candidate
that would generate the detected Cherenkov ring with an accuracy on the mm scale, hence ensuring the desired charge resolution.
An alternative approach would have been to improve the light collection uniformity by coupling a short light guide (15-20\,mm in length) to each tube, covering a square photon collection area. 
The light collection efficiency would have been improved from 50\% to 90\%, and the homogeneity increased by a factor of 4.
However, this option has been discarded because a direct photon counting wouldn't have been possible without correcting the induced Cherenkov ring deformation.
Additionally, the light guides cost and weight were unacceptable for this embarked experiment.

\section{Mechanics}
\subsection{Description}
\begin{figure}[th]
\begin{center}
\includegraphics[angle=-90,width=15cm]{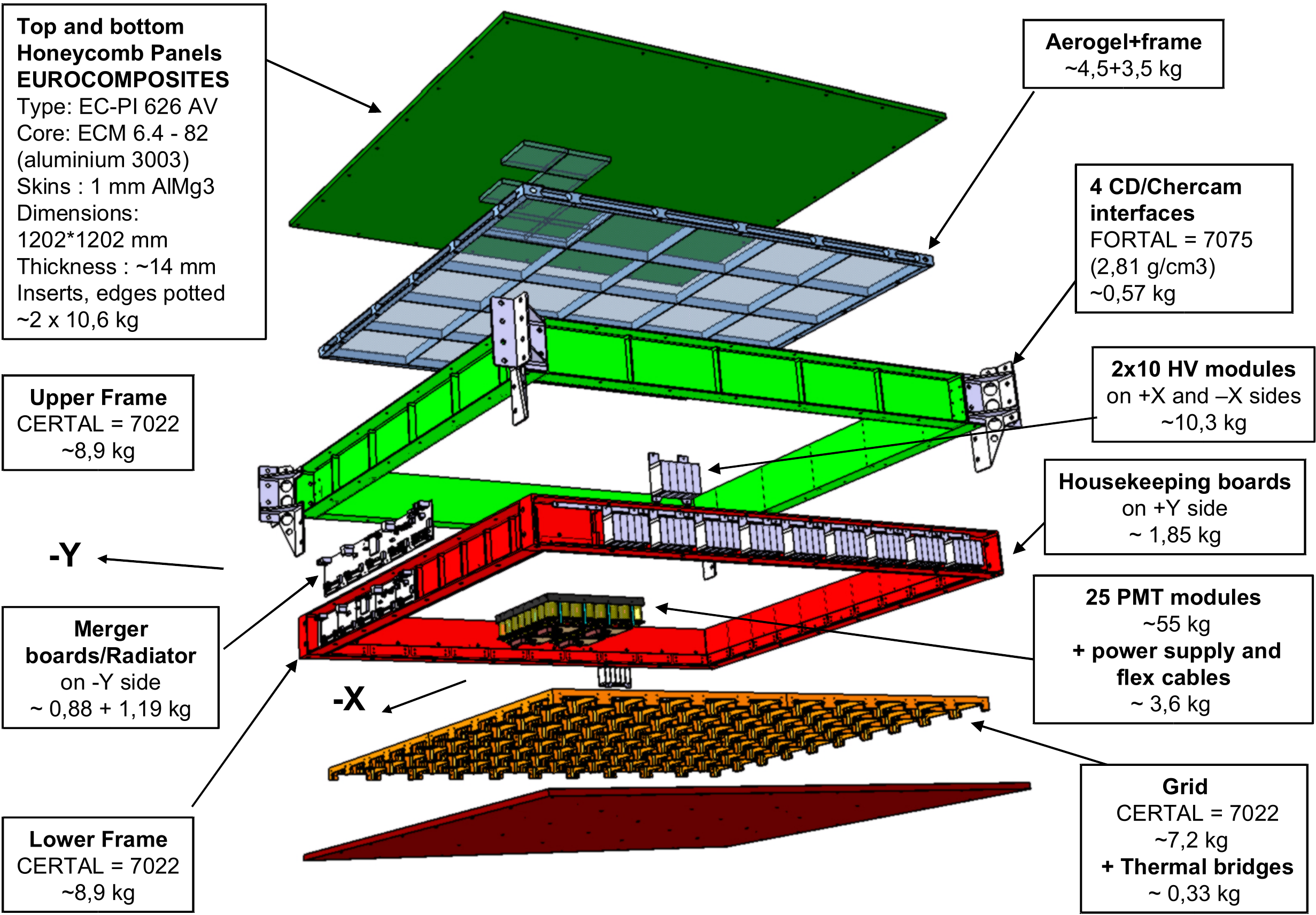}
\caption{Exploded view of CHERCAM showing the main components.}
\label{ChercamEclate}
\end{center}
\end{figure}

\begin{figure}[th]
\begin{center}
\includegraphics[width=8cm]{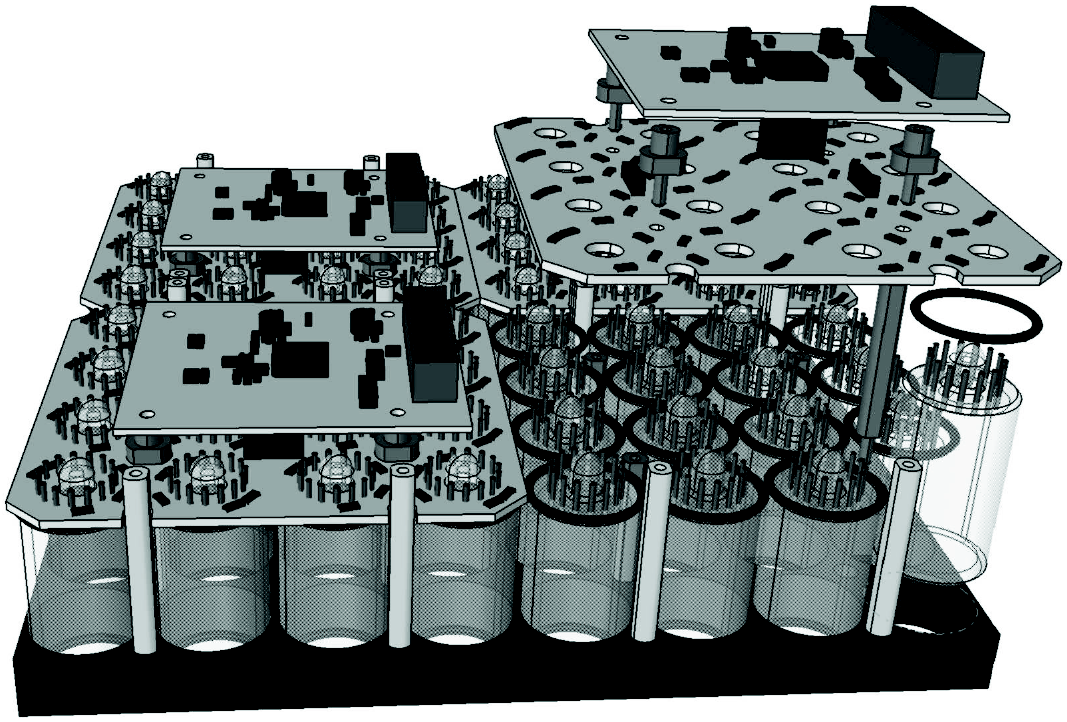}
\caption{Exploded CAD view of one CHERCAM module, composed of $2 \times 2$ submodules of 16 PMTs inserted in a 15\,mm thick Ertalyte\textregistered \ block, each with 64 bore holes of $25.7_{-0}^{+0.02}$\,mm diameter. The PMTs of one submodule are padded with rubber O-rings at their readout end and hold in place by the HV distribution board. Each submodule is clamped down using spacer screws which attach to 4 spacer rods. The spacer screws are in turn used to attach the front end readout board to the HV distribution board. This way the module becomes one stable unit of 64 PMTs ensuring their appropriate positioning in the array with a 27.5\,mm pitch at a tolerance of $\pm$0.02\,mm. The 3 rows of 4 long spacers, located along the submodule edges, are used to attach the module to the supporting grid.}
\label{ExplodedSubmodule}
\end{center}
\end{figure}

CHERCAM's main body is made of two superimposed certal\textregistered \ (aluminum alloy) frames. The upper frame is dedicated to the radiator and photon drift volume. The lower frame is housing the PMT array, High Voltage (HV) dividers, and associated front-end electronic readout. The external dimensions of CHERCAM are $\rm 120.2 \times 120.2\ cm^2$, while the internal fiducial area is $\rm 110 \times 110\ cm^2$ (figure \ref{ChercamEclate}). The upper and lower ends of the two superimposed frames are closed by 1.4\,cm thick lids of aluminum honeycomb sandwiched between two 1\,mm foils of $\rm AlMg_3$.
The upper lid holds the radiator plane (described in section \ref{AerogelSection}).
The lower frame contains a support grid on which 25 modules are fixed in a square pattern of $5 \times 5$ modules. The grid itself is fixed on the frame body. 
%
Figure~\ref{ExplodedSubmodule} shows an exploded CAD view of one CHERCAM module, composed of $2 \times 2$ submodules of 16 PMTs inserted in a 15\,mm thick Ertalyte\textregistered \ block, each with 64 bore holes of $25.7_{-0}^{+0.02}$\,mm diameter. The PMTs of one submodule are padded with rubber O-rings at their readout end and hold in place by the HV distribution board. Each submodule is clamped down using spacer screws which attach to 4 spacer rods. The spacer screws are in turn used to attach the front end readout board to the HV distribution board. This way the module becomes one stable unit of 64 PMTs ensuring their appropriate positioning in the array with a 27.5\,mm pitch at a tolerance of $\pm$0.02\,mm. The 3 rows of 4 long spacers, located along the submodule edges, are used to attach the module to the supporting grid.
This architecture ensures the appropriate PMT positioning in the array with a 27.5\,mm pitch.

The lower honeycomb lid closes the bottom of the lower frame and contributes to the mechanical rigidity of the structure. The sides of the lower frame are hosting the 20 high voltage modules and first level electronics: `Merger', `HV control', and `housekeeping' boards. In order to avoid light leaks, all cable feedthroughs were filled with black silicon sealing. Blind vents were drilled in the counter frames to allow constant pressure balance during the ascending and descending phases of the payload (large pressure variations between 1\,mb and 5\,mb).
Finally, the technical choices allowed CHERCAM's total weight to be kept below 130\,kg.
 

\subsection{Thermal study}
\begin{figure}[th]
\begin{center}
\includegraphics[width=10cm]{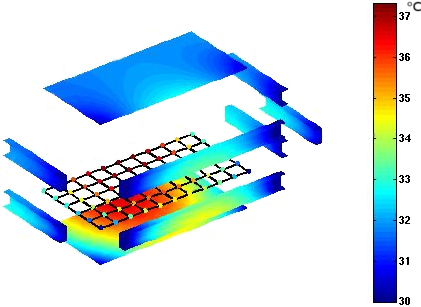}
\caption{Temperature map across CHERCAM from global scale simulation, assuming all PMTs supplied with 950\,V with a converter efficiency of 0.45 and the heat conducted through the holding bracket (contact resistance 570\,$\rm W/m^2K$)\textbf{ and limit condition fixed at 30\textcelsius}.}
\label{CHercamTemp}
\end{center}
\end{figure}

\begin{figure}[th]
\begin{center}
\includegraphics[width=0.8\textwidth]{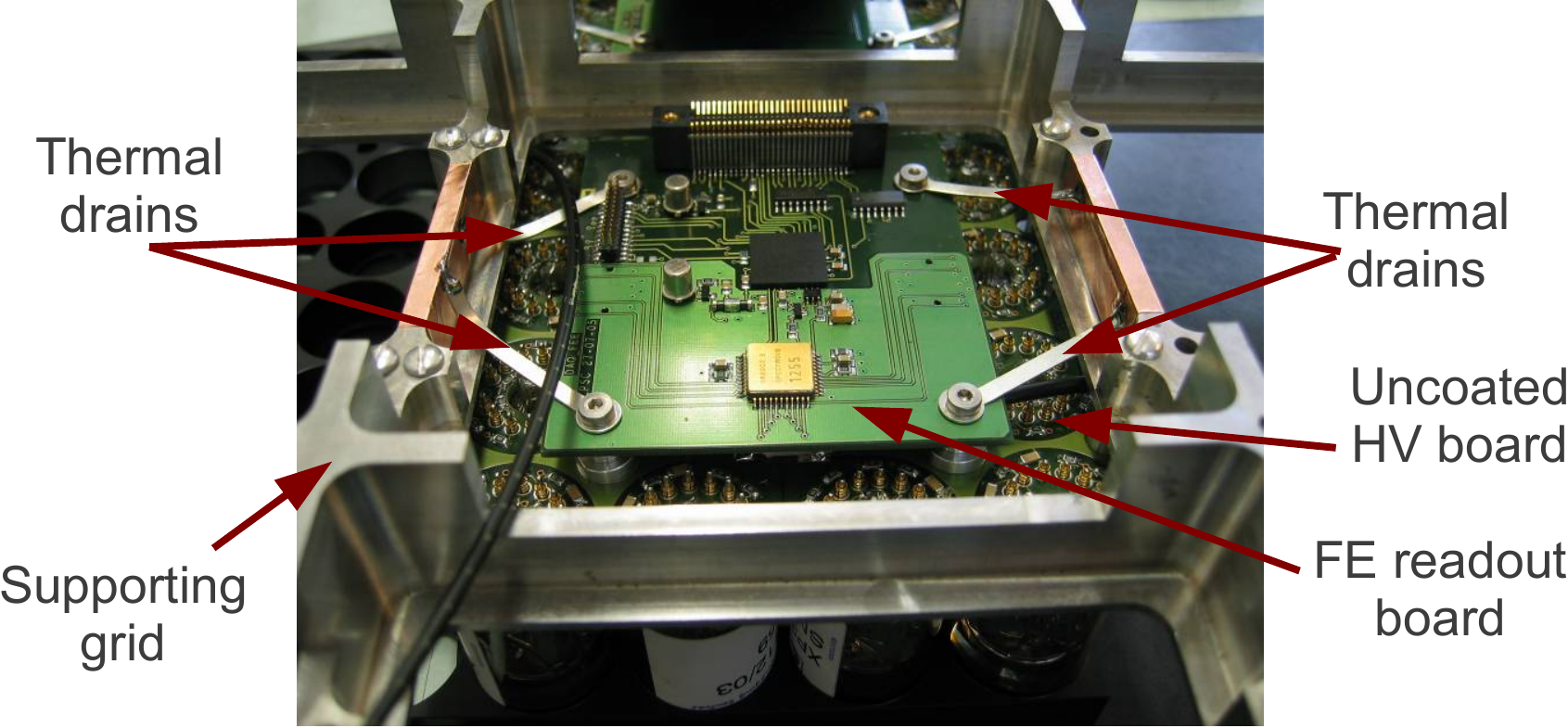}
\caption{Submodule mounted in the grid with 4 thermal drains installed.}
\label{ThermalDrain}
\end{center}
\end{figure}

As CHERCAM is operated at low pressure, air thermal conduction and convection are negligible.
A passive cooling method was considered and implemented in order to keep design simplicity and to remain compatible with the CREAM experiment.
CREAM is dissipating its power through heatpipes connected to radiative black painted panels located outside of the instrument casing, where the temperature is about -100\textcelsius. The thermal analysis was performed both at local and global scale.\\
The global scale simulation was used to determine the temperature gradient across the mechanical structure as a function of different PMT operational conditions and mounting options. One of the simulated scenarios is presented in figure~\ref{CHercamTemp}, corresponding (at the time of the design) to a conservative case where the converter efficiency was estimated to be 0.45 (instead of 0.55), the HV set to 950\,V for all PMT (real operational average of 758\,V) and the limit condition of 30\textcelsius \  was used instead of the really experienced 25\textcelsius. 
Due to the ineffective heat conduction in the structure, the highest temperature spot is located in the middle of the grid.\\
The local simulation scale was used to put realistic constraints on the electronic power consumption and to determine the best mounting options in order to decrease as much as possible the central temperature, prejudicial to the PMTs. To reduce the effect of heat dissipation by the front-end electronics in the center of each submodule the following two measures were taken:
\begin{itemize}
\item The total resistance of the HV divider chain of each PMT was increased from 90\,M$\Omega$, as recommended by the manufacturer, to 130\,M$\Omega$,
thereby reducing the heat dissipation of the 16 divider chains of one submodule from 100\,mW to 69\,mW;
\item The thermal conductance between the submodule electronics and the grid by was increased by adding thermal drains (see figure~\ref{ThermalDrain}), 
which provided a known and controlled heat conduction path guarantying the maximum temperature elevation while adding only 1.5\,grams of weight to the submodule.
\end{itemize}
%
It was estimated that with these two measures implemented the residual temperature increase during flight operation would be only  0.4\textcelsius \ for the HV distribution board and 3.4\textcelsius \ for the Front-End (FE) readout board.
\subsection{Structural analysis}
The mechanical stresses in the CHERCAM structure have been investigated using the ``Generative Structural Analysis'' toolbox from Catia V5\copyright\  \cite{CATIA}. Several loading scenarios to the structure have been computed: airplane transport, a symmetric and eccentric parachute snatch, shocks up to accelerations of 10\,g, etc.
To facilitate the computations, CHERCAM main components and interfaces were modeled as homogeneous linear elastic and isotropic solids. Consequently, the honeycomb panels were replaced by equivalent homogeneous plates. Tetrahedral elements were then used to mesh the structure. Other components, e.g. the electronic boards, were replaced by point-like or homogeneous masses. The mechanical interfaces of CHERCAM are mounted to the Instrument Support Station (ISS) brackets. Each interface incorporates an additional bore hole in order to carry the Cherenkov Detector (CD) of the CREAM experiment. The CD weight (60\,kg) is regarded in the model as equally distributed over the four bore holes mentioned.

The ends of the CHERCAM interfaces were assumed to be perfectly bonded to the ends of the ISS brackets. Additionally, the ISS body and brackets were assumed to be perfectly rigid. Simple boundary conditions were thus assigned, i.e. the ends of the CHERCAM interfaces were clamped. In practice, the structural components are fastened together using bolts or rivets. 
Therefore, the main contact surfaces between the distinct parts located inside the instrument were assumed to be perfectly bonded. 
The contacts between CHERCAM interfaces and CHERCAM frame were however modeled with frictionless unilateral contact conditions. 
In order to take into account the pre-tensions, a bolt tightening connection was defined for each bolt connecting these structures. Consequently, the stresses in each bolt could be computed in a conservative way. 

\begin{figure}[th]
\begin{center}
\includegraphics[width=0.9\textwidth]{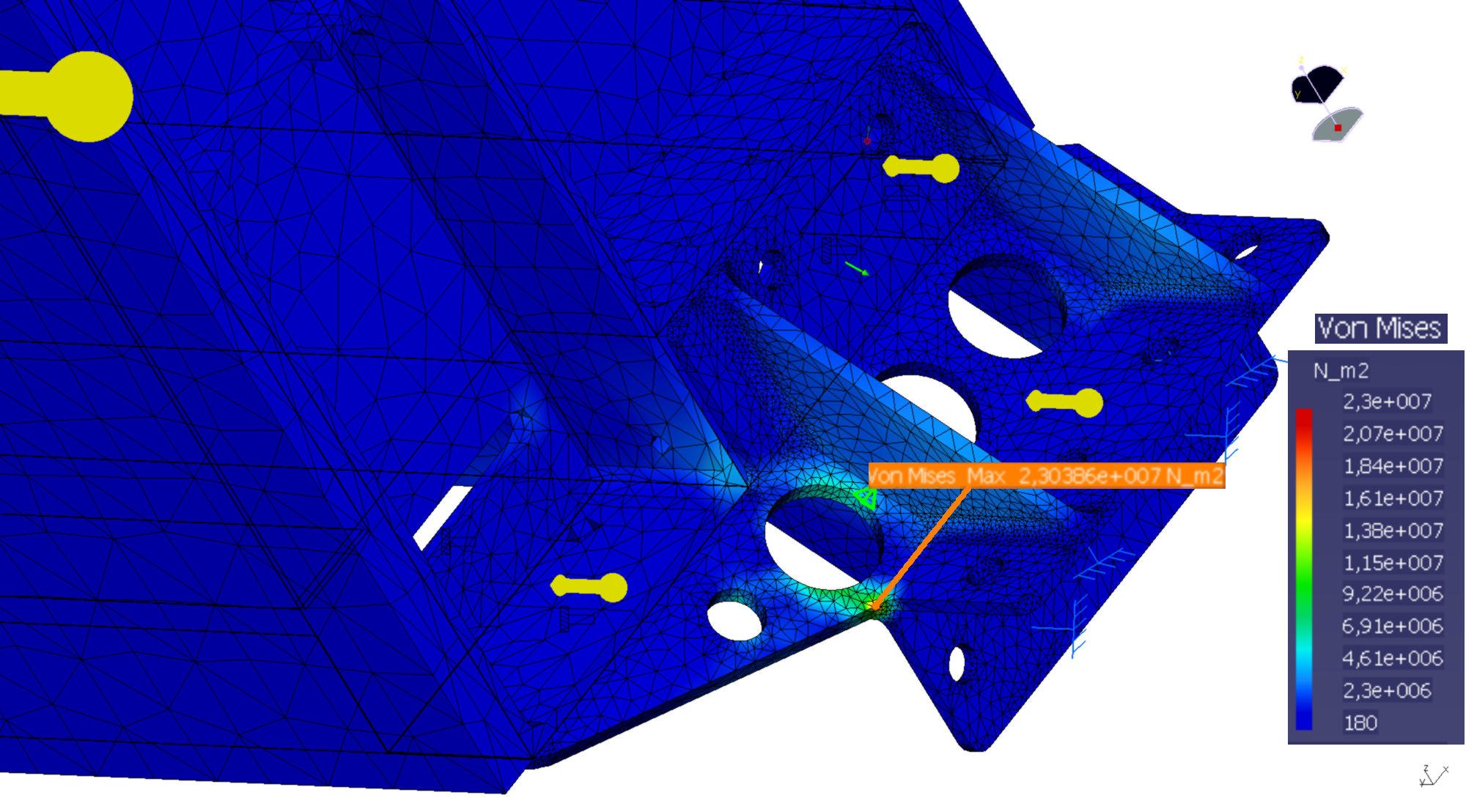}
\caption{Result of the structural analysis when the instrument is subjected to a horizontal acceleration of 5\,g (yellow arrows). Zoom on the interface. The maximum Von Mises stress is reached in a fillet.}
\label{JNP_accel_5G}
\end{center}
\end{figure}

The deformation of CHERCAM when subjected to its own weight was estimated first. The magnitude of the maximum displacement found in the central part did not exceed 0.04\,mm (i.e. of the same order of magnitude as the geometrical tolerances). Thus, it indicates that the deformation of the photon detector plane can be neglected in the GEANT4 simulations. Equivalent Von Mises stress fields were then computed for various extreme loading conditions. The maximum stress is reached in the interface for a vertical acceleration of 10\,g and is estimated to be less than 25\,MPa. Figure~\ref{JNP_accel_5G} presents a zoom on the interface showing the result for a side acceleration of 5\,g. The maximum Von Mises stress is reached in a fillet of the interface, but remains very small with respect to the yield stress of the chosen 7075 alloy. It was found that the main fasteners and the overall structure were far from yielding in any of the considered cases.

Unfortunately, the actual boundary conditions that should be applied on the brackets are much more complex than assumed above. The interfaces can not be considered as simply clamped. The kinematics of the contact zone between the interfaces and the brackets is linked to the rigidities of the ISS and of the other detectors. Consequently, a refined finite element analysis of CHERCAM was performed by the NASA engineering department \cite{EYO}, after being included in the CREAM instrument. A simplified CAD model and a detailed description of the masses of the non modeled parts was provided. It was shown that all the CSBF\footnote{Columbia Science Balloon Facility} structural requirements, in particular concerning CHERCAM and the fasteners, were fulfilled. From a practical point of view, no mechanical damage to the instrument following various transports, operations and landings was observed.

\section{Electronics}

\begin{figure}[th]
\begin{center}
\includegraphics[angle=-90,width=10cm]{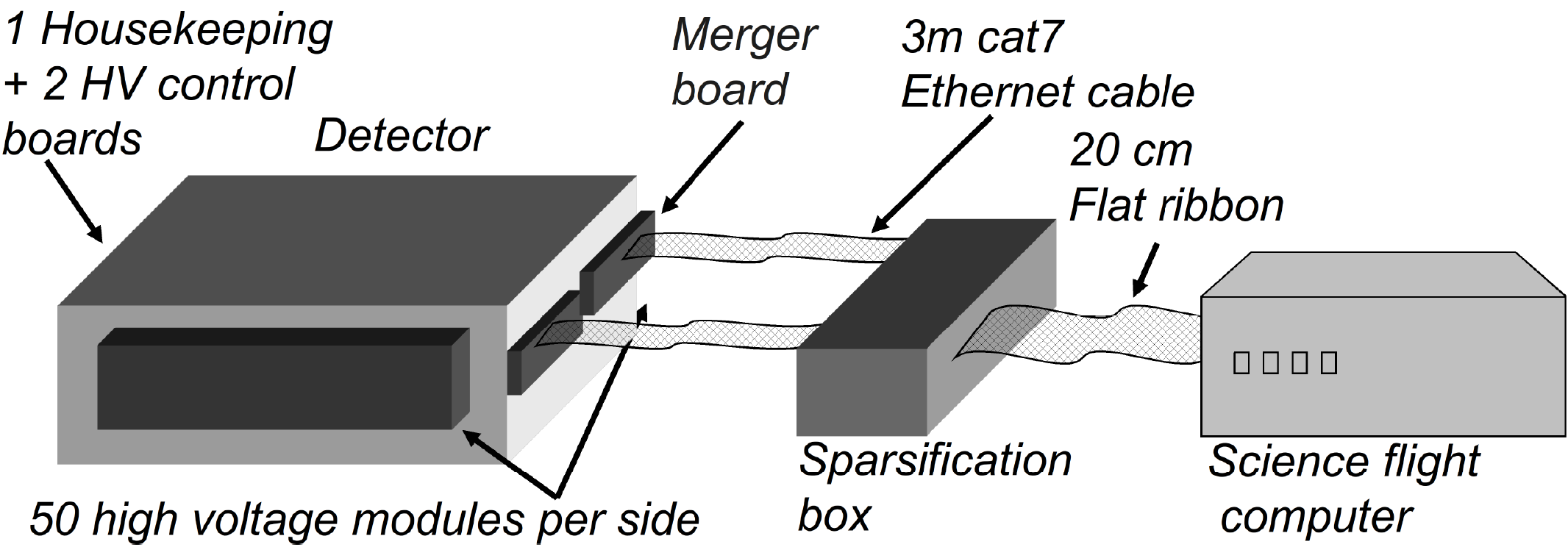}
\caption{Overview of the CHERCAM electronics. The low power supply box is not represented.}
\label{SystemOverview}
\end{center}
\end{figure}

The detector electronic structure encompasses three levels of processing, each of them is located at different places in the instrument, as shown on figure~\ref{SystemOverview}: the front-end readout and merging electronics are located in the CHERCAM body just underneath the PMT plane, where as the Sparsification and the Science Flight Computer (SFC) are situated on the data acquisition (DAQ) desk at the bottom of the instrument~\cite{Ahn}. 

Since CHERCAM is operated at around 40\,km altitude and powered by solar panels, special care was taken in designing the electronics in order to meet the critical points of power consumption ($<$\,60\,W in total), dissipation budget, and radiation tolerance. These constraints drove the choice to \textit{ACTEL proASIC plus\textregistered \ } FPGA, which are flash based. At the time of the design, using CREME96 simulations and extrapolating the results from \cite{Wang,Wang2} to the 0.22\,\textmu m FPGA technology, indicated that in a 100 days, less than 3 Single Event Latchup (SEL)  and 5 Single Event Upset (SEU)  should be experienced per flight for 105 FPGAs. In addition, all power supplies were equipped with fast tripping current safeties in order to avoid destructive latch up. 

\subsection{The HV insulation issue}
Carrying High Voltages in a low atmospheric pressure environment is a very challenging undertaking. In this particular case, the residual pressure in flight conditions is around 5\,mbars, i.e. at the Paschen minimum where the risk of electrical breakdown is most critical \cite{Bloess}. As a result, all HV connections had to be carefully insulated, PC boards coated, and PMT sockets volumes potted with appropriate insulation materials. This was a source of strong complication in the design and in the construction procedure.

\subsection{Front end electronics}
\begin{figure}[th]
\begin{center}
\includegraphics[width=7cm]{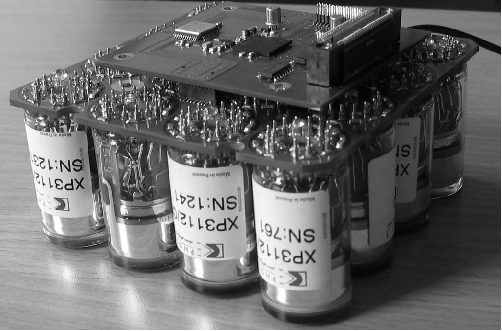}
\caption{Picture of a submodule with the 16 PMTs mounted directly on the HV distribution board (rubber O-ring not inserted), topped with a readout front end card. The interface connector with the FLEX PCB shows on the top right. The PMT bases are not potted and high voltage board is not covered with insulator.}
\label{SousModule}
\end{center}
\end{figure}

The readout of the 1,600 PMTs and first level data conditioning is performed by 100 readout sets, each of the latter containing a high voltage distribution board and a Front-End (FE) readout board (see figure~\ref{SousModule}). 

\subsubsection{The HV distribution board}
\label{HVdistribsec}
The HV distribution board fulfills two functions.
Firstly, the HV distribution board provides the mechanical support for the submodule, with the PMTs directly soldered into the cut-out sockets on the printed circuit board (PCB).
The protruding PMT dynode pins on the readout side of the HV distribution board as well as the volume on the HV divider side between the PCB and the PMT base had to be potted with resin material to insulate all surfaces subjected to HV potential.
The insulation had to be fault-free for the system to operate safely at the low atmospheric pressure.
Practically, this was achieved by placing rubber O-ring seals between the base of the PMT tubes and the PCB at the time of mounting, to prevent the resin insulator from leaking between the PMT tubes during the potting.
The still liquid resin was then carefully syringed by hand from the back of the PCB through the central opening into the cavern created on the PMT side. Great care was taken to leave no air bubbles which would compromise the insulation.
Then the PMT pins on the readout side as well as the resistor divider bridge were covered by resin.
The mechanical requirements of this process required the PMTs to be sorted and grouped in classes by tube length, a parameter which is not very well controlled during PMT production and can vary by a few millimeters between tubes. A length variation of $\pm$0.3\,mm was tolerated.
%

Secondly, it feeds the high voltage from a single HV channel to the 16 voltage divider chains and distributes the potentials to the PMT dynodes. For board layout consideration, it was not considered to use a common voltage divider.
Due to the counting rate being expected as low as only a few Hz for the entire CHERCAM detection plane, and that each PMT involved (up to 10) detects most of the time a few photons only, it was possible to increase the total resistance of the divider chains to 130\,M$\Omega$ in order to minimize the power consumption. The divider ratios between the PMT's dynodes were maintained as recommended by the manufacturer.  This modification was done without degrading the PMTs linearity.
It was validated experimentally on a dedicated test setup using a LED as a light source. The amount of light produced as well as the trigger rate were tested according to the CREAM experiment foreseen conditions.


Within each tube length class, the PMTs were grouped by supply voltage yielding a single photon electron gain of 5 ADC channels, with a tolerance of $\pm$10\,V. This corresponds to a gain tolerance of $\pm$8\%.
The sorting was done by experimentally determining, for each PMT, the single photoelecton gain as a function of the high voltage supply. 
The method used to determine the PMT gain in ADC count (G) relied on the fact that the  number of detected photoelectron follows a poissonian distribution having a mean value \textmu.
Each PMT gain was then obtained from its spectrum average value (G\textmu) and from the mean number of photoelectron (\textmu), computed with the following relation $\rm \mu=\ln \frac{N_{tot}}{N_{ped}}$ where $\rm N_{tot}$ is the total number of counts and $\rm N_{ped}$ is the number of counts in the pedestal.


\subsubsection{FE readout electronics}
\label{FEreadoutsec}
\begin{figure}[th]
\begin{center}
\includegraphics[angle=-90,width=0.9\textwidth]{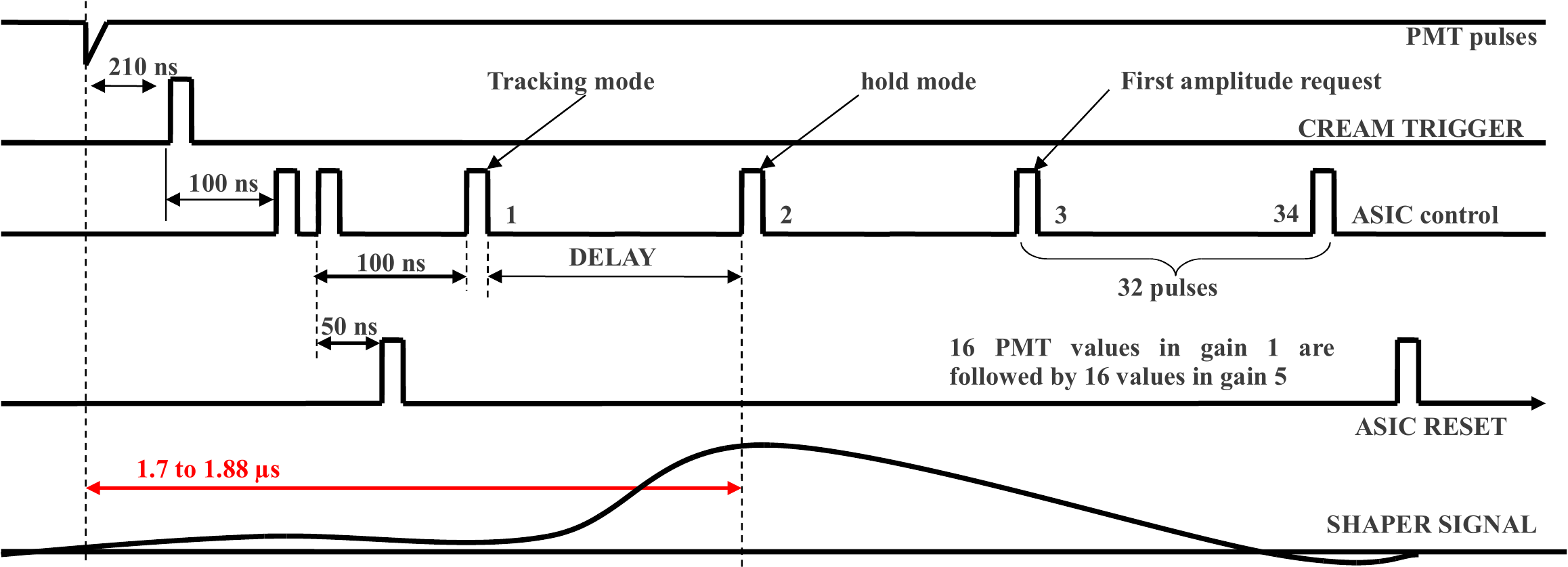}
\caption{Timing diagram showing a typical ASIC control sequence. The first pulse puts the ASIC in the tracking mode and second one in the hold mode. The 32 following pulses are used to switch between the 16 channels first provided in gain 1, then in gain 5. The 2 pulses shown before the first ASIC reset pulse are to fix a known design defect.}
\label{AsicTiming}
\end{center}
\end{figure}

The FE readout electronics were based on the same ASIC as previously developed for the AMS RICH \cite{LGM}. This 16 channel ASIC features for each channel: a charge sensitive preamplifier followed by a shaper having a typical peaking time of 1.7\,\textmu s to 1.88\,\textmu s, 2 amplifiers (1x, 5x) in order to optimize the ADC range and match the required PMT dynamic range, and a track\&hold circuitry. The ASIC also features a multiplexer which is used to select the channel and gain to be monitored. The track\&hold circuitry and the multiplexer are externally controlled by 34 25\,ns-wide digital pulses. The first pulse puts the ASIC in the tracking mode and second one in the hold mode. The 32 following pulses are used to switch between the 16 channels first provided in gain 1, then in gain 5. A typical control timing sequence is provided figure~\ref{AsicTiming}. 

\begin{figure}[th]
\begin{center}
\includegraphics[width=0.7\textwidth]{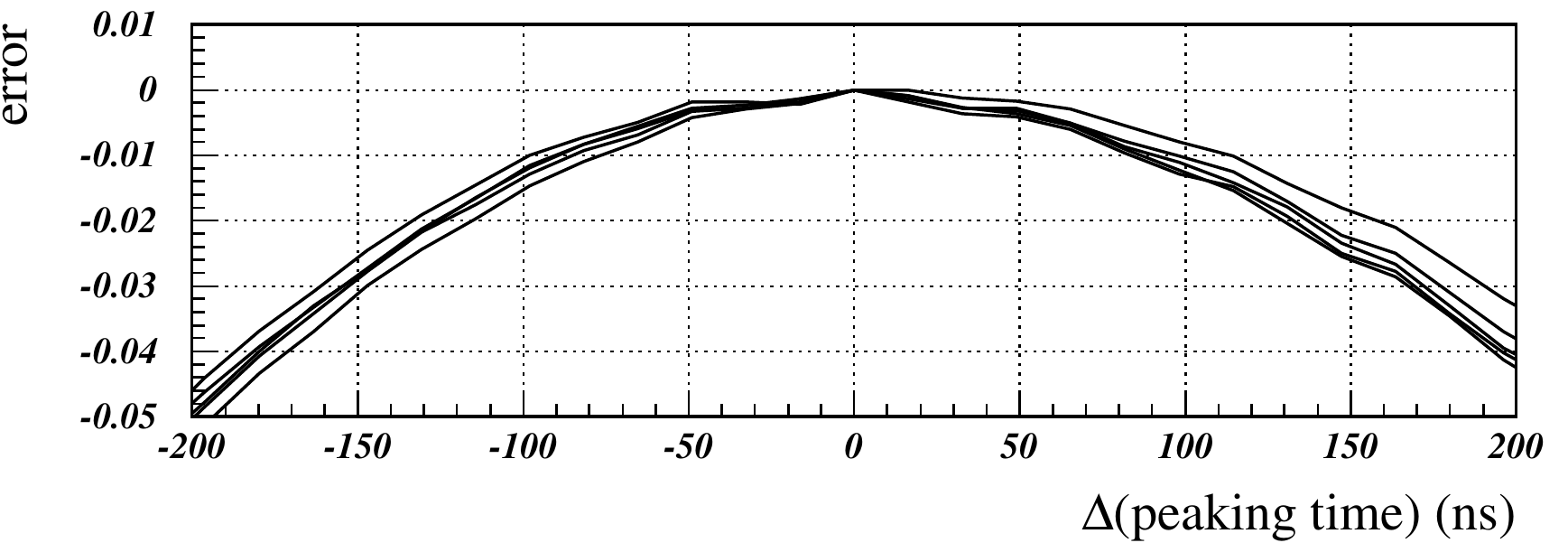}
\caption{Amplitude error plotted as a function of the peaking time variation. Less than 1\% of error can be achieved by sampling the shaper output signal within $\pm$75\,ns of the optimal value. 5 channels of a single ASIC are plotted and it can be seen that the inter-channel dispersion is particularly negligible in the plateau.}
\label{PeakingTimeAsic}
\end{center}
\end{figure}

The control of the track\&hold circuitry, which correspond to the first pulse of the sequence, must be timed within less of $\pm$75\,ns of the optimal peaking time value in order to compensate the mean peaking time dispersion between ASICs, the inter-channel dispersion within an ASIC being negligible. Figure~\ref{PeakingTimeAsic} shows the amplitude error as a function of the sampling time.

\begin{figure}[th]
\begin{center}
\includegraphics[width=0.7\textwidth]{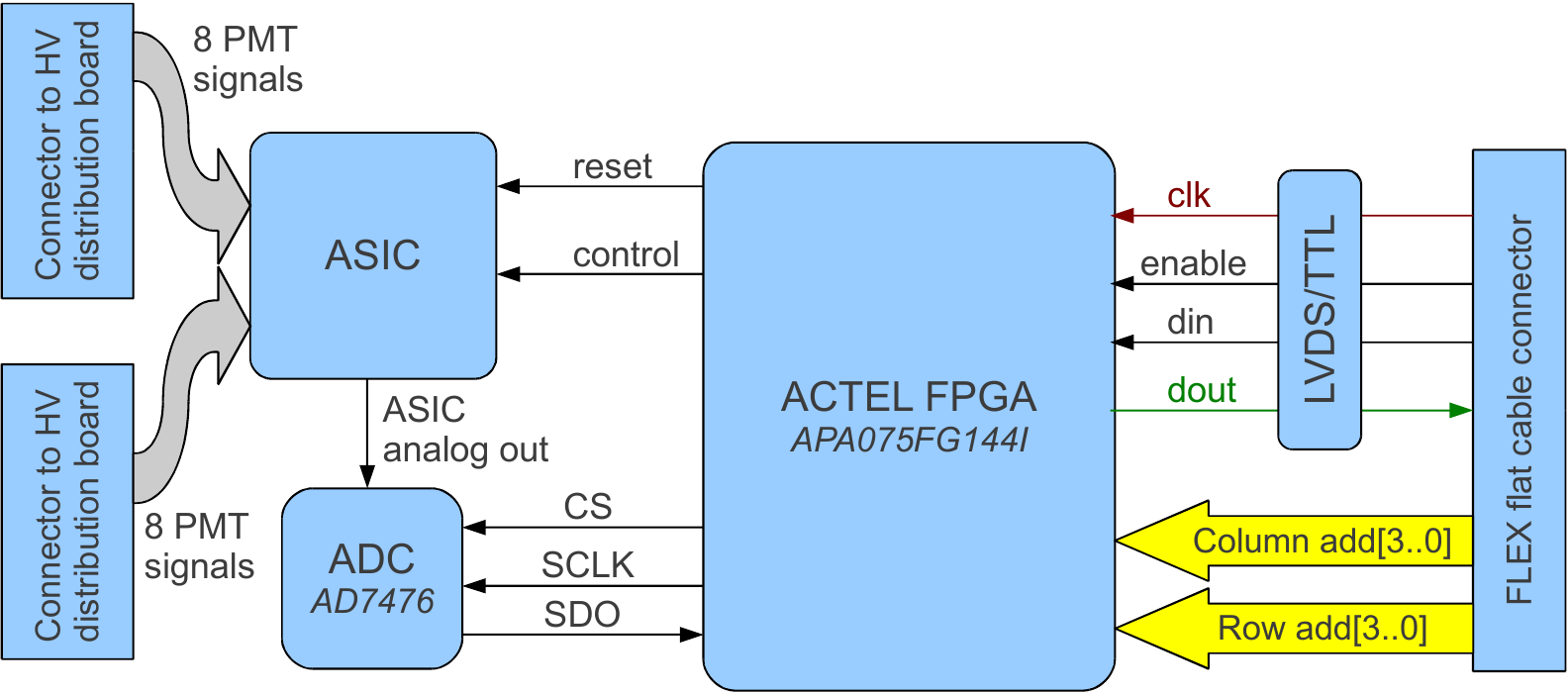}
\caption{Block diagram of the FE readout board. Each board features one flash based FPGA, one ADC and one ASIC. The power supplies, communication layer and the submodule address are provided by the FLEX PCB.}
\label{DaqFeeBlock}
\end{center}
\end{figure}

\begin{figure}[th]
\begin{center}
\includegraphics[angle=-90,width=0.9\textwidth]{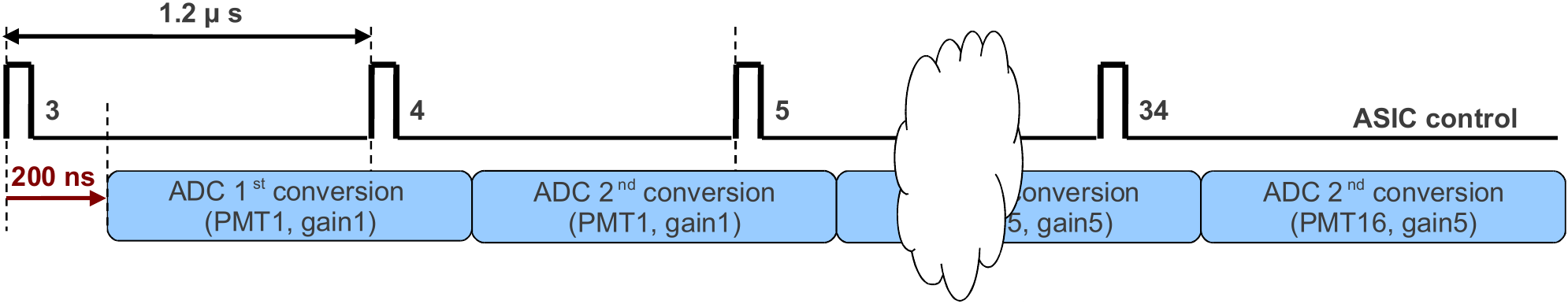}
\caption{Timing diagram showing the interleaving of ASIC track\&hold control and the ADC conversion. The first 200\,ns corresponds to the ASIC output time. The total conversion time of the 32 voltages is 38.6\,\textmu s.}
\label{AsicADCInterleave}
\end{center}
\end{figure}

The fast external trigger is provided by either the scintillator hodoscope (TCD) or the calorimeter (CAL)~\cite{Ahn}.
For better flexibility and reduced connectivity, a flash based Actel FPGA (APA075\-FG144I) running at the system clock frequency of 40\,MHz was implemented on each readout board, with the  main purpose of controlling the ADC in charge of digitizing the ASIC information upon trigger reception, see figure~\ref{DaqFeeBlock}.
To achieve this function, it awaits the reception of the trigger commands, controls the track and hold circuitry with a preprogrammed delay value expressed in steps of 25\,ns (roughly equal to the peaking time minus the fast external trigger propagation time) and digitizes all channels in the 2 gains (see figure~\ref{AsicTiming}). The chosen ADC is a commercially available AD7476 (version A) which was qualified and tested for radiation tolerance by the AMS experience. It has a 12 bit resolution for an input dynamic range of 3.3\,V. The track\&hold  control and the ADC readout are interleaved, as shown in figure~\ref{AsicADCInterleave}, in order to minimize the total time required. The 32 voltages values are successively converted in 38.6\,\textmu s.

The second purpose of the FPGA is to enable the communication with the DAQ electronics via a custom serial protocol using LVDS signals at a clock speed of 40\,MHz transmitted via a FLEX PCB. In addition of providing a communication medium, the FLEX PCB also carries the power supply voltages and individual submodule addressing. The coordinate address of each submodule is fixed by its position on the FLEX PCB and the Merger port it is attached to. This type of interconnection minimizes the mass at which passing particles can scatter and produce secondary particles.

Communication with the next readout stage, the Merger board, is limited to 16 32-bit data words per submodule and event. This choice was made to reduce the transmission time and thus to use less power (details about clock commutation in section~\ref{MergerDescription}). In the normal data acquisition mode an algorithm in the FPGA selects which of the signals, gain x1 or gain x5, is transmitted for each PMT. Whenever a gain x1 digitized PMT value is below a fixed value of 512, the gain x5 is used. In two calibration modes the gain selection can be set to be the same for all PMT using either gain.

\subsubsection{Dynamic range discussion}

Knowing that for Z=1, depending on the crossing angle 5 to 10 photons are produced and that a typical Cherenkov ring hits 10 PMTs of the detector plane, this yields a mean number of photoelectrons (in the ring) per PMT of 0.5 to 1. For Z=26 that yields 338 to 676 photons detected per PMT. As explained  in section~\ref{HVdistribsec}, during the sorting process, the single photoelectron gain was chosen to be at 5 ADC count in gain x1.
 This choice was motivated by the necessity to have a sufficient separation between pedestal peak ($\sigma$=5) and single photoelectron peak in gain x5 (G=25). Assuming a PMT linear response, the largest signal to detect (with one sigma margin) is supposed to be 3438, which fits a 12 bit ADC  dynamic range. 
\begin{figure}[th]
\begin{center}
\includegraphics[width=0.5\textwidth]{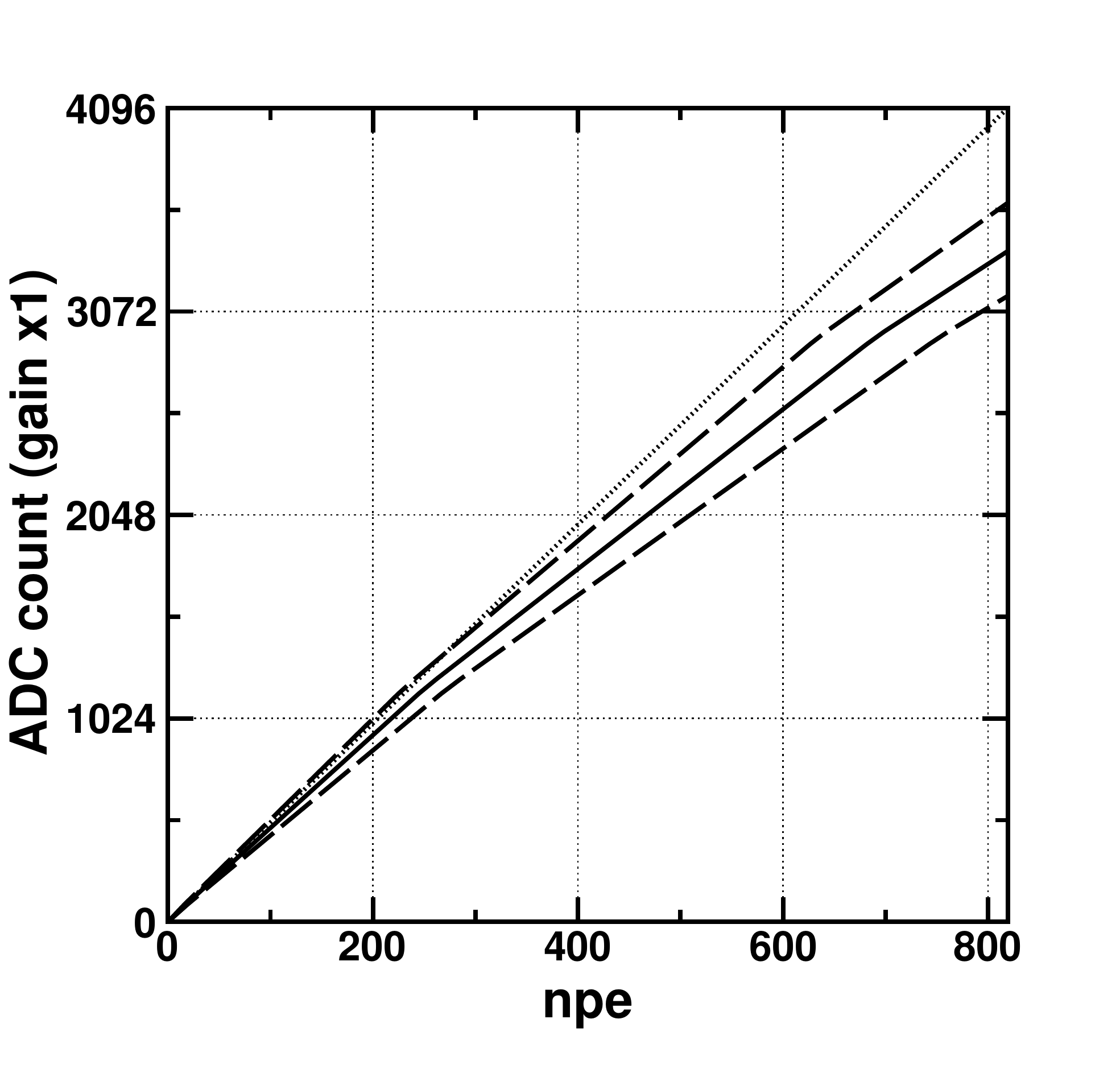}
\caption{Plot of the PMT response, expressed in ADC channels at gain x1, as a function of the number of detected photoelectron (npe). The dotted line represents the ideal PMT response, the solid line represents the real response with a single photoelectron gain of 5 and the dashed line shows the envelope corresponding to the tolerance of $\pm$8\%.}
\label{Gain1}
\end{center}
\end{figure}

In practice the combined contribution of the single photoelectron gain fluctuation ($\pm$8\%) due to the sorting tolerance and of the pedestal (up to 200 ADC counts) are reducing the available dynamic range. Both of these effects are, however, mitigated by the non-linearity of the PMT response.
Figure~\ref{Gain1} shows the PMT response, expressed in ADC channels at gain x1, as a function of the number of detected photoelectron. It can be seen that even with the largest single photoelectron gain there is still a dynamic range margin of $\sim$1000 ADC counts for 676 photons (Z=26), which is enough to compensate the pedestal contribution.

\subsection{Merger board description}
\label{MergerDescription}
The Merger boards are primarily used to interface the 100 FE submodules to the Sparsification board. FLEX PCB are used as a communication medium between the FE electronics and the Merger. The communication control lines going from the Merger boards to the FE electronics are using multidropped (10 sinks) LVDS signals while the readout line from FE boards toward the Merger board is point to point. 

The control commands are, with one exception detailed later, issued by the Sparsification board and indiscriminately repeated to the FE electronics by the Merger boards. The FE clock signal is activated only during the readout or command sequences. This significantly reduces the power consumption as the counting rate is low and most of the DAQ related power is drawn by the digital part of the FE readout electronics.

The control communication is based on a serial synchronous link with a data enable signal. The enable signal is required for the control sequences to be accepted the electronics. The clock used in this link is also used as the system clock for the FE and Merger boards and is running at 40\,MHz. Two types of control commands are implemented, 2-bytes commands and 1-byte commands. With 2-byte commands the FE electronics FPGAs parameters which hold the delay values of the ASIC channels can be updated. 2-byte commands also are used to set the intensity of the LED pulser and to trigger the LED pulser in calibration mode. The trigger and all other control commands, e.g. to set the readout mode or to reset, use the 1-byte format.

\begin{figure}[th]
\begin{center}
\includegraphics[width=11cm]{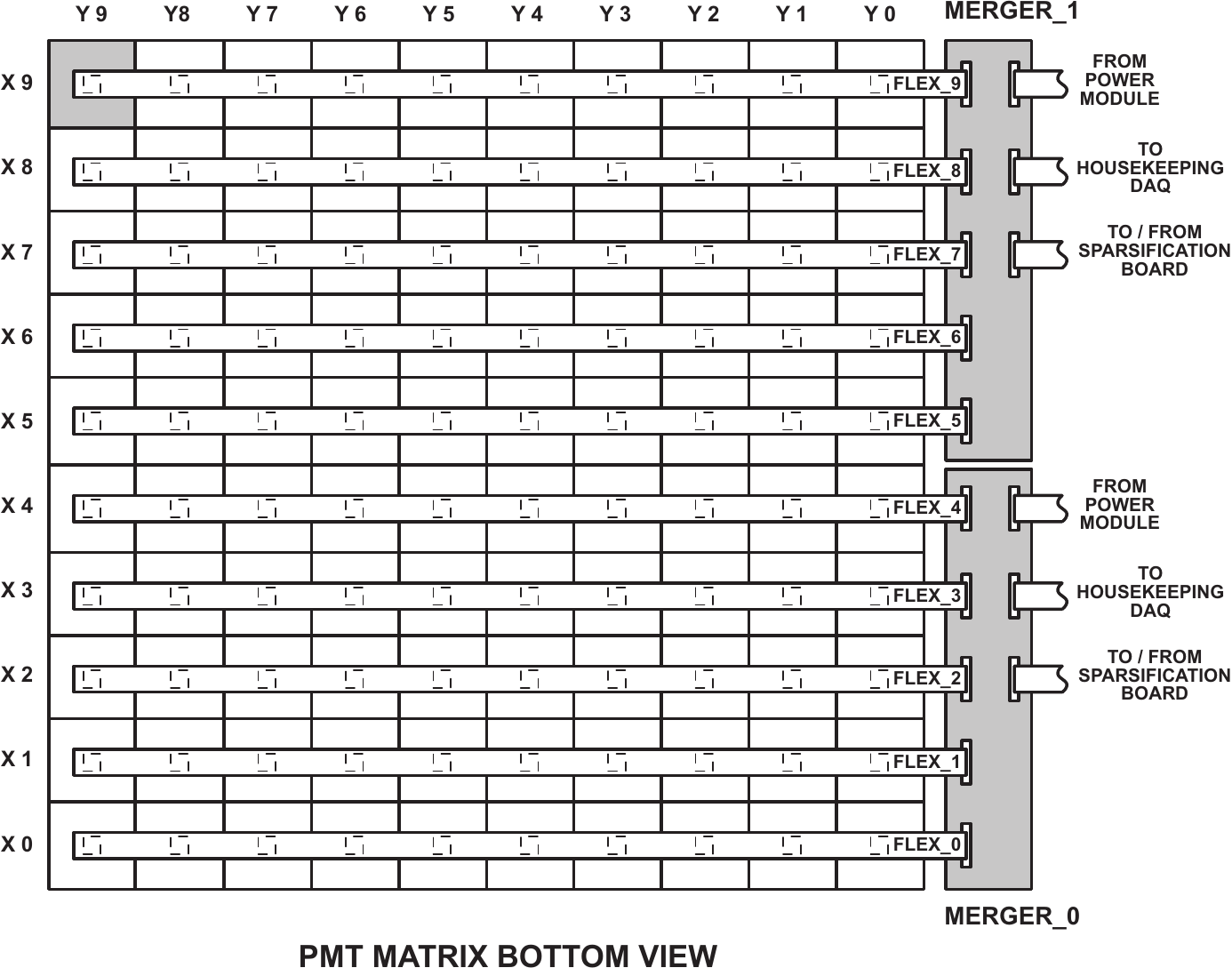}
\caption{Overview of the submodules interconnection and link with the merger board. FE submodules coordinate addresses as a function of the FLEX port and the merger port connection are shown.}
\label{SubModuleAddressing}
\end{center}
\end{figure}

After a fixed time, when even the most delayed signal is guaranteed to have been converted in the submodules, each Merger board 'polls' row by row the FE boards for data transmission. The readout geometry matches that of the PMT array with 10 parallel readout lines of 10 PMT submodules (see figure~\ref{SubModuleAddressing}).

 Since each dedicated readout line on the FLEX PCB is routed directly from each submodule connector to the Merger, the propagation time is a function of the submodule position. The submodules of a row are equally separated on the FLEX PCB, which is 1.1\,m long in total (composed of two parts of 55\,cm). Consequently, in order to avoid any sampling issue and to provide reasonable timing margin, the simultaneous data transmission from the 10 FE electronics of each row to the Merger board is done asynchronously at $\rm 1/8^{th}$ of the system clock rate.
The data received, which are deserialized during reception, are multiplexed and forwarded to the Sparsification board via the LVDS data link \cite{ChannelLink}.

\begin{figure}[th]
\begin{center}
\includegraphics[width=10cm]{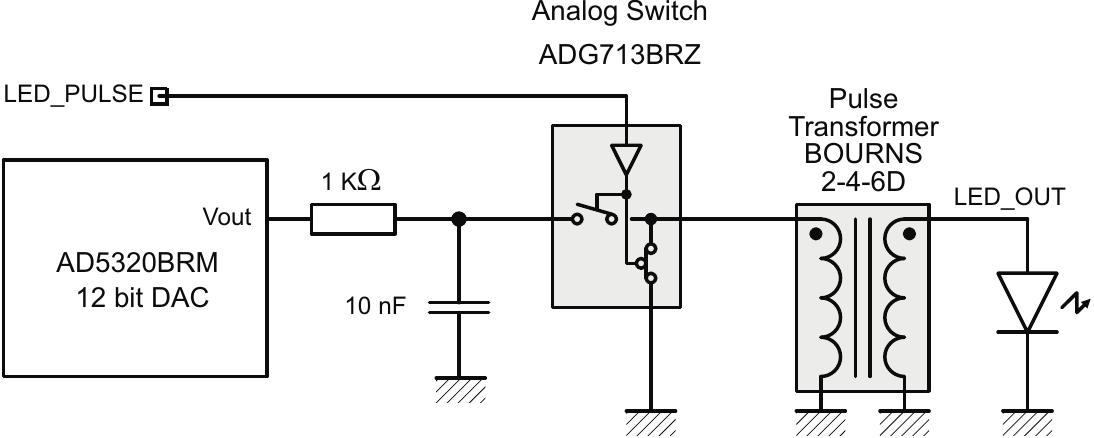}
\caption{Block diagram of the LED control system. The light intensity of the LED pulse is controlled via a DAC and the activation pulse, whose  fixed duration is 25\,ns, is used to stimulate the LED via a pulse transformer  having an impedance ratio of four.}
\label{LedSynop}
\end{center}
\end{figure}

The second purpose of the merger is to operate the LED in calibration mode, which allows PMT gain monitoring during the flight. In this mode,
 instead of solely propagating the trigger command to the submodules, a LED pulse is fired. Consequently the delay values used by the FE FPGA to control the ASIC track\&hold are priorly adjusted accordingly to the increased delay in the LED calibration  mode. 
The LED control system, which is available on both Merger boards, is shown in figure~\ref{LedSynop}. The light intensity of the LED pulse is controlled via a DAC and the activation pulse, whose  fixed duration is 25\,ns, is used to stimulate the LED via a pulse transformer having an impedance ratio of four.

\begin{figure}[th]
\begin{center}
\includegraphics[width=7cm]{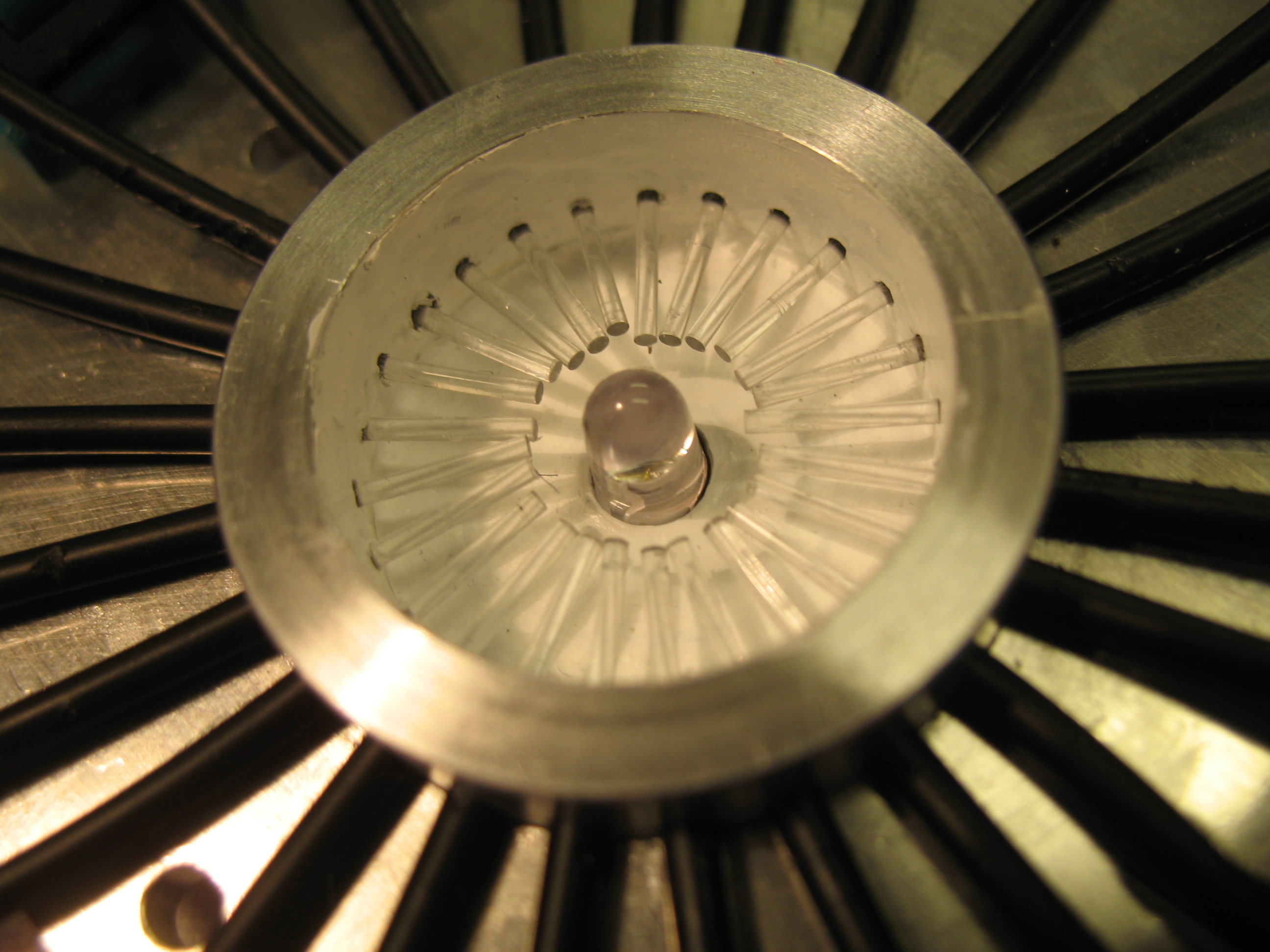}
\caption{Picture of the LED distribution system. The LED light flash is distributed to 25 fibers.}
\label{LEDspider}
\end{center}
\end{figure}

Only one LED is used to illuminate the entire CHERCAM detector plane. The light produced by a single LED is transferred by 25 optical fibers by a distribution system (shown in figure~\ref{LEDspider}) directly mounted on only one of the Merger boards. The 25 fibers are emerging in the center of the Ertalyte\textregistered\  blocks and pointing toward the aerogel which is used as a reflector and diffuser to illuminate the detector plane. 

\subsection{Data acquisition}
\label{DAQ}

\begin{figure}[th]
\begin{center}
\includegraphics[angle=-90,width=13cm]{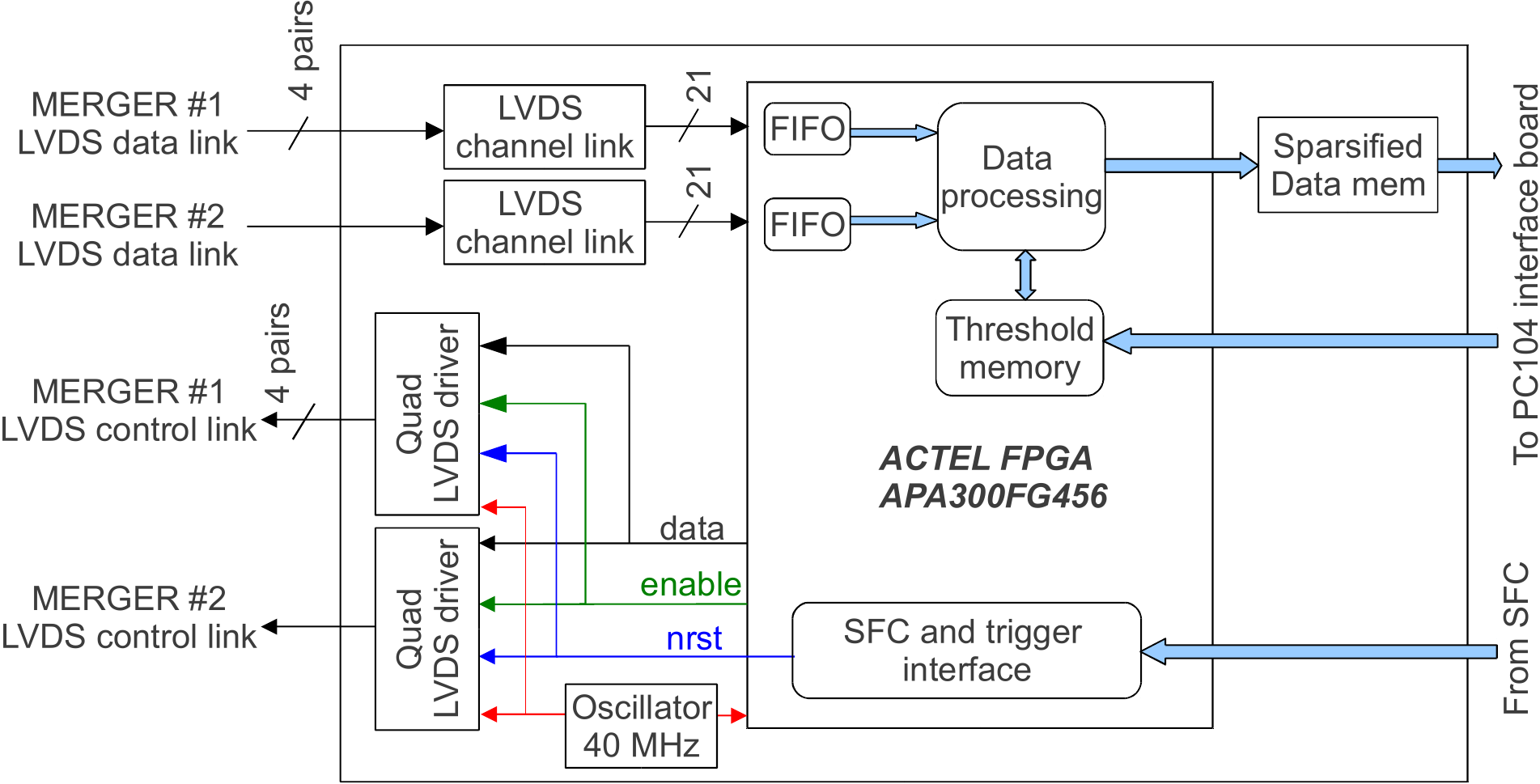}
\caption{Block diagram of the sparsification board.}
\label{sparsification}
\end{center}
\end{figure}

The Sparsification board (figure~\ref{sparsification}) serves several purposes. 

At first, it interfaces the trigger system of the CREAM experiment to the CHERCAM system.
If a particle interaction passes the experiment trigger conditions, the Sparsification board receives a trigger signal with a delay of 210\,ns, as well as a 32-bit event number.
The trigger is then encoded as a command and transfered to the FE electronics in order to initiate the readout.
With the front-end deterministically responding to the trigger, i.e. in real-time, the data is received via the two LVDS data links from 800\,ns after forwarding the trigger via the two LVDS control links. The data link are using the 'channel link' circuits \cite{ChannelLink} which provides out of the box high speed serializing/deserializing solution. The data rates involved demands the uses of CAT7 patch cable used as serial data links. Those feature four twisted pair data lines per cable with a guaranteed minimum skew. The pairs are individually and globally shielded.

In data taking mode, the Actel (APA300BG456) FPGA on the Sparsification board concatenates the data from the two Merger boards, performs the zero-suppression and writes the remaining data to the sparsified data memory (its format is shown in table~\ref{DataBuffer}).
\begin{table}[th]
\begin{center}
\begin{tabular}{|c|c|c|c|c|c|c|c|}
\hline
address/bit field & 31$\rightarrow$26 & 25 & 24 & 23$\rightarrow$20 & 19$\rightarrow$16 &15$\rightarrow$12 & 11$\rightarrow$0 \\
\hline
\hline
0 & \multicolumn{7}{|c|}{Event number} \\
\hline
1 & \multicolumn{5}{|c|}{status word} & \multicolumn{2}{|c|}{Event size}\\
\hline
2$\rightarrow$2+event size & unused & data valid & Gain & Y & X & channel & ADC value\\
\hline
\end{tabular}
\end{center}
\caption{Organization of the 32 bit words sparsified memory.}
\label{DataBuffer} 
\end{table}

For the zero-suppression, individual thresholds are used for each channel. It is performed when a channel value is received in gain x5 and is below threshold, gain x1 values are systematically kept. These thresholds are derived from signal spectra obtained in 'pedestal' calibration, details in section~\ref{Calib}.

The global readout and control system access the Sparsification memories and register via a PC104 parallel interface, a typical read or write cycle takes between 70\,ns and 100\,ns. The sparsified data memory is used in a double buffering mode and a control register shared with the global readout system provides handshaking. 
As a result of the zero-suppression e.g. at an event rate of 2\,Hz and a typical event size of 20 photon hits a bandwidth of only 1.4\,kbits/s is required, which fits the allocated bandwidth. The data are also stored on flash disk by the CREAM SFC. 

A further role of the Sparsification board is to receive commands from the Science Flight Computer (SFC) and to forward them to the Merger boards via the LVDS control links. The serial synchronous link protocol used is the same as described in section \ref{MergerDescription}. The same kind of patch cable are used to transport the signals.

\subsection{Power supplies}
The power supplies are a key issue in balloon-borne experiments in terms of reliability and conversion efficiency. They are supposed to build all the necessary  voltage from the main 28\,V power line with a minimum of electromagnetic/radio frequency interferences (EMI/RFI).

\subsubsection{Low voltage}
The Low Voltage Power Supply box (LVPS) is made of several industrial grade DC-DC converters stabilized by steep low pass filters. Each output of the LVPS is equipped with a solid state switch controlled remotely by the SFC and locally by a fast trip circuit. On each power line the fast trip circuit monitors the current drawn and deactivates the output if it exceeds 110\% of the nominal current. These limits are chosen to prevent destructive single event latch-up.

\subsubsection{High voltage}
\label{HVPSsec}
\begin{figure}[th]
\begin{center}
\includegraphics[width=0.5\textwidth]{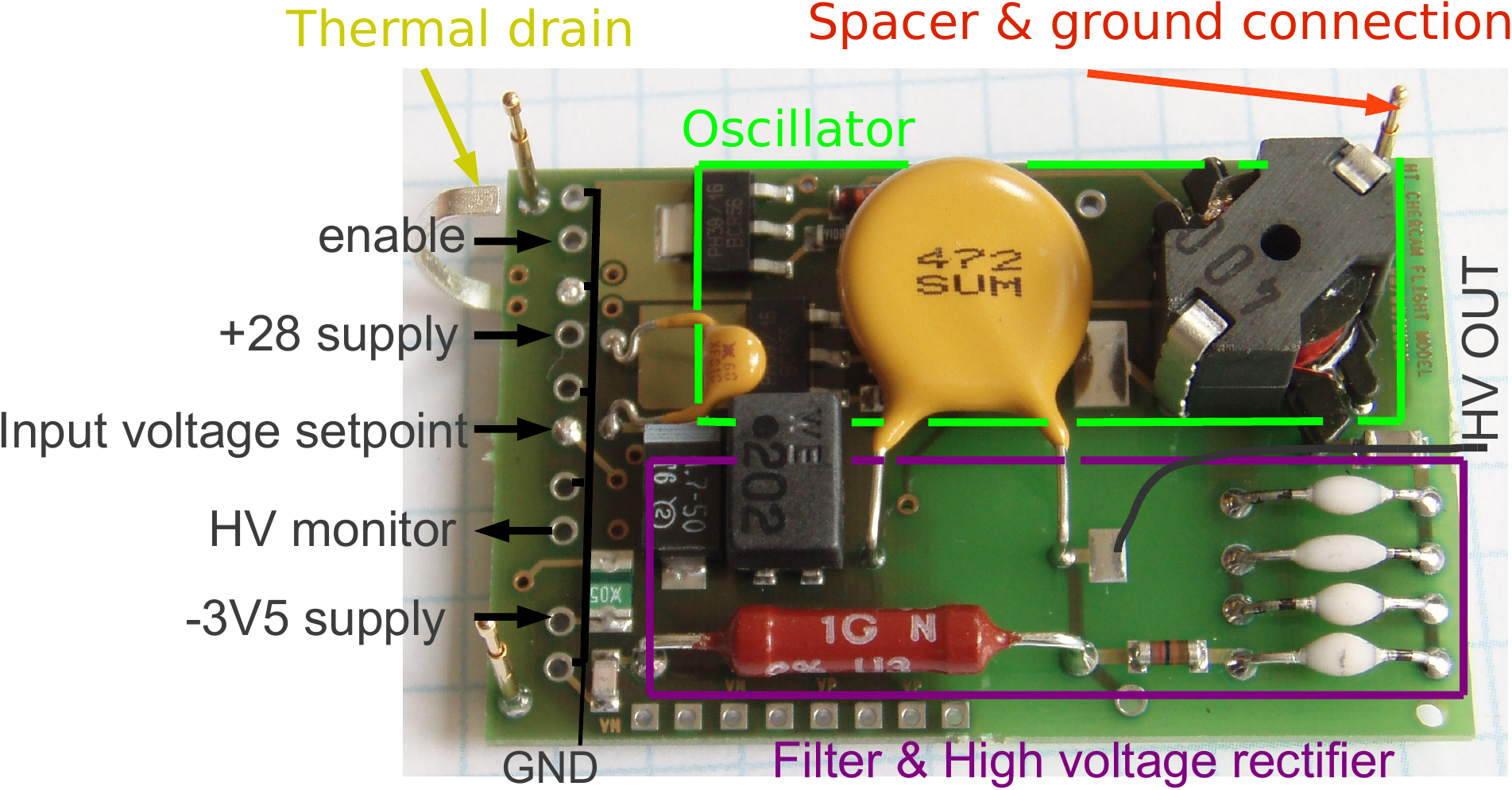}
\caption{Picture of the high voltage board, actual size is 52\,mm \texttimes \ 33\,mm.}
\label{HVpic}
\end{center}
\end{figure}

The 100 HV Power Supplies (HVPS) were designed to be operated at an altitude of 40\,km, where the Paschen minimum is reached. 
Each of the supplies powers the 16 PMTs of a submodule. The voltage can be adjusted in the range 0\,V to -1400\,V with a resolution of 340\,mV.
The design is built around an auto-oscillating circuit yielding a sinusoidal waveform. It is based on a homemade transformer and two bipolar transistors \cite{medale}\footnote{\label{fn:secret}The technical details were not disclosed by the developers}). This oscillator is followed by a two-stages Cockroft Walton high voltage amplifier (figure~\ref{HVpic}). The output voltage is regulated by a proportional corrector comparing a fraction of the output high voltage with the setpoint. One of the major benefit of this design is to lower the conducted and radiated noise as the $\sim$120\,kHz oscillation is sinusoidal. The unit is supplied with a main 28\,V power line and a -3.5\,V power supply line used for the regulation electronics. A HV monitoring signal, having a ratio of 1/1000 is provided for remote unit surveillance.
With the equivalent load of one submodule (16 PMTs), the maximum power efficiency of 70\% is reached at the high voltage of -1350\,V. At this setting the power consumption, measured on the 28\,V input line, is less than 10\,mA per HVPS. At nominal conditions of 750\,V the power efficiency is around 55\% and the power consumption about 125\,mW per HV channel.

A thin coaxial high voltage  Reynolds\textregistered\  cable\footnote{AWG26, rated 18\,KV, sheath in Fluorinated Ethylene Propylene} was used for HV distribution within the detector. The cable ends were treated with a TETRA-ETCH\textregistered\ compound to obtain a good grip for the Solithane S113\textregistered \  resin.

\begin{figure}[th]
\begin{center}
\includegraphics[width=0.4\textwidth]{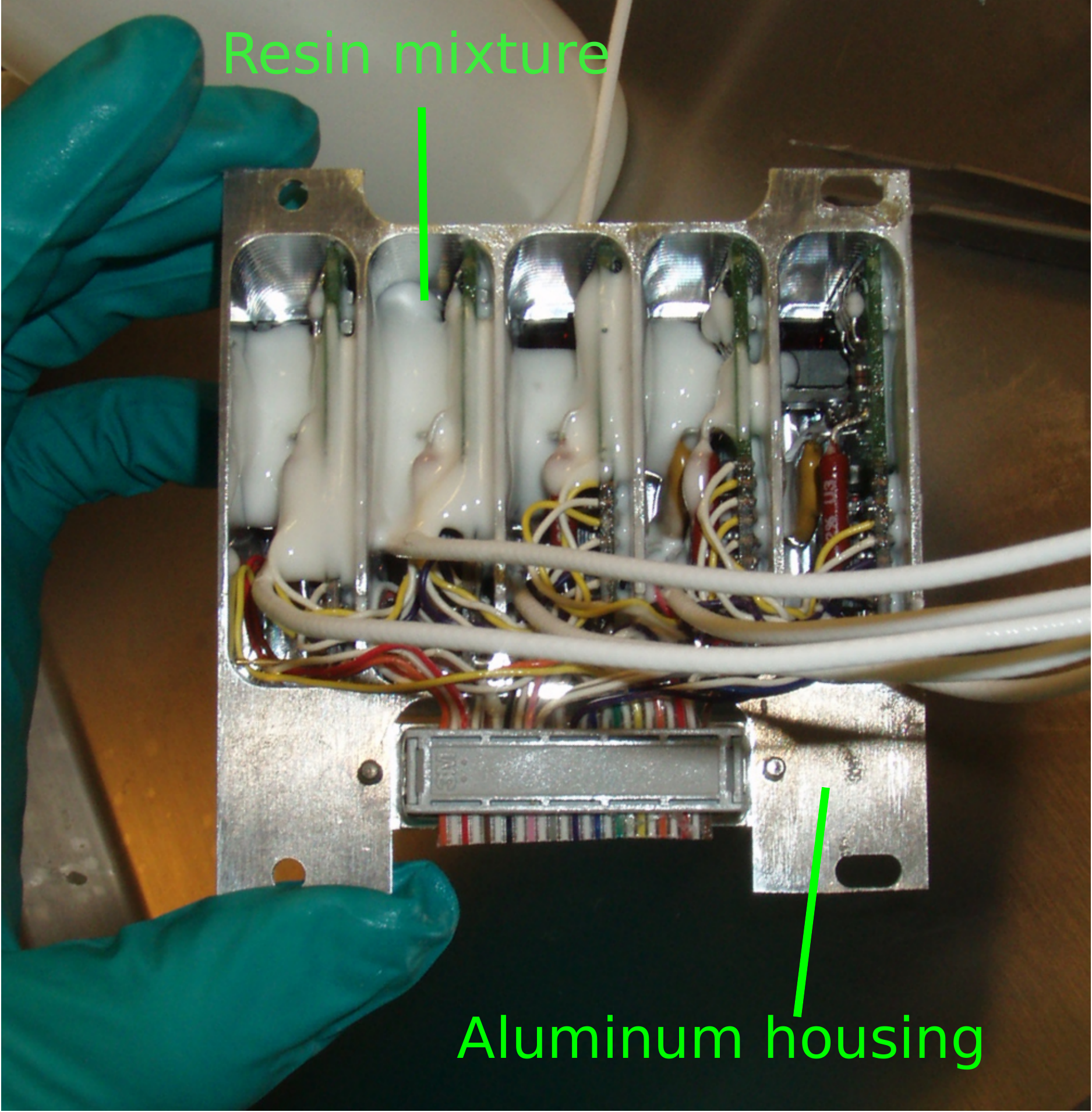}
\caption{Picture of one module containing 5 HV boards in an intermediate stage of the potting operation.}
\label{HVmodule}
\end{center}
\end{figure}

Groups of five HVPS boards each were mounted in a common aluminum alloy housing, shown in figure~\ref{HVmodule}, which shielded the oscillators noise to a level acceptable to the analog readout electronics nearby.
To improve the radiative heat dissipation the Al-boxes were painted black on the outside.
For potting of the HVPS boards a highly efficient insulating mixture of silica and Solithane S113 R resin was used  \cite{LeComte}\textsuperscript{\ref{fn:secret}}.

\subsection{Slow control}
\label{SlowControl}
The FE and Merger boards are configured via command words issued by the Sparsification board, as described in section \ref{MergerDescription} and \ref{DAQ}. The LVPS boards are controlled by the Science Flight Computer, see section \ref{HVPSsec}. 

The 100 HVPS boards are controlled by two dedicated, identical HV control boards located on one side of the lower frame of CHERCAM.  Each HV control board uses one ACTEL\textregistered\  FPGA (APA300BG456I) and 50 12-bit DACs (AD5328) to individually steer the HVPS boards. The FPGA controls the enable signals and configures the DACs, which in turn provide analog output in the range of 0\,V to 6\,V. 25-wire ribbon cables are used to transmit in parallel these signals as well as the power lines and to retrieve the HV monitoring voltage.

In addition a housekeeping board is placed between the HV control boards. It interfaces the $\sim$120 signals monitored in the CHERCAM system, multiplexes the gathered signals and transmits them back to the global CREAM housekeeping system via a 25-wire ribbon cable. The multiplexing is controlled directly by the SFC, which sends an address before each digitization. All signals are converted once per second.
Among the monitored quantities are the high and low voltages, their currents and twelve temperatures. The integrated-circuit temperature sensors (LM35) were placed at locations previously determined as the most critical by the simulation, i.e. on the hottest points of the electronics and mechanical structure and on one central PMT. 

\section{In-flight detector calibration and monitoring}
\label{Calib}
In order to allow good instrumental performance, CHERCAM was designed to be
calibrated regularly (once per hour) during flight using the different DAQ
modes provided (see section~\ref{DAQ}). 

\begin{figure}[th]
\begin{center}
\includegraphics[width=0.48\textwidth]{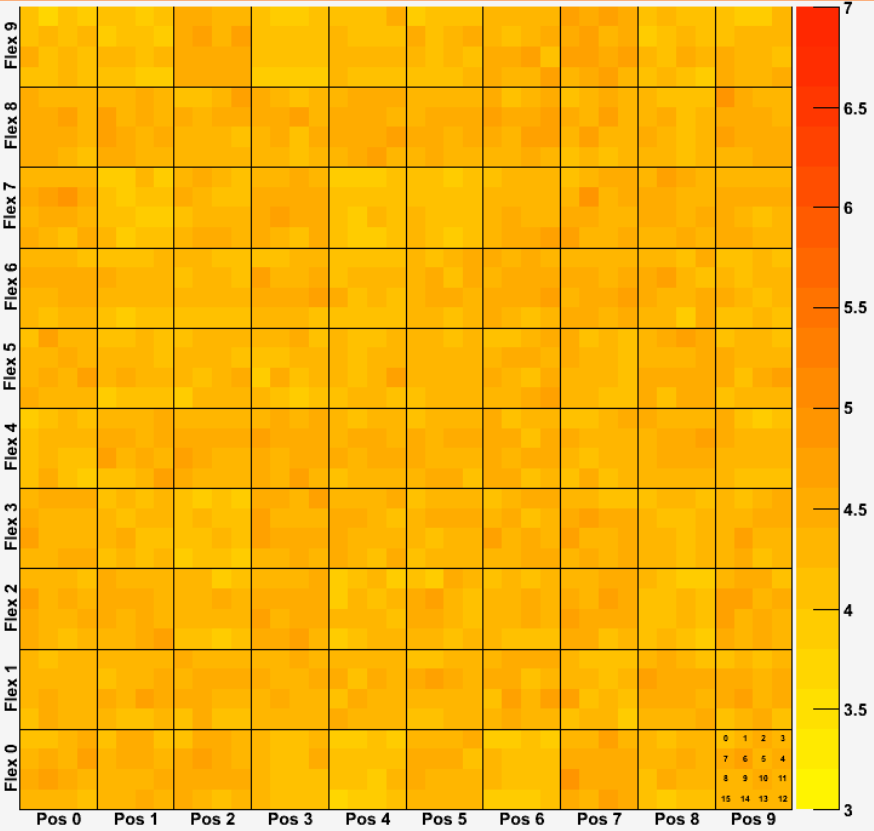}
\hspace{5pt}
\includegraphics[width=0.48\textwidth]{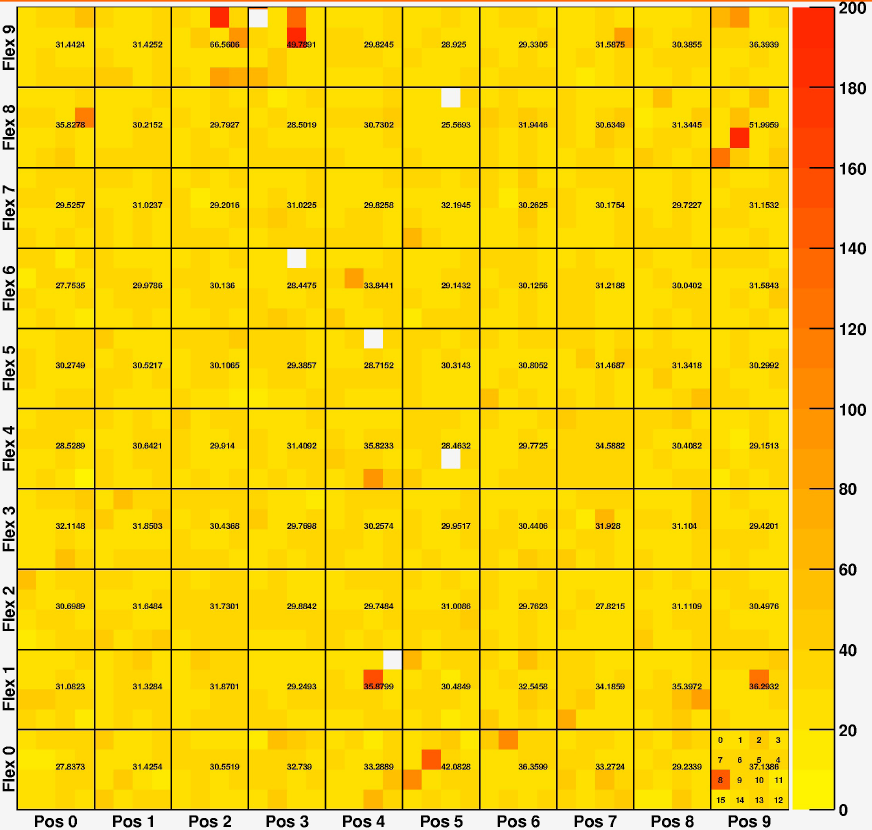}
\caption{Typical results of the calibration modes (gain x5). Left picture shows
the width of the pedestal in ADC count (i.e. the noise figure), right picture
shows the gain in ADC count. The 6 white square correspond to unavailable
channels. The plots are expressed as a function of the channel position.
\label{PlotCalib}}
\end{center}
\end{figure}

In a first calibration run, lasting approximately 5 to 10\,s (at a 200\,Hz rate), the pedestals are measured using the gain x5.
From this run the offsets, the electronic noise and the slow thermal drifts can be determined.
Monitoring the pedestals in one gain only is sufficient and saves time, as they dominantly depend on the characteristics of the preamplifier inside the ASIC (on the FE boards) which is located before the gain selectable amplifier.
Using each channel spectrum, individual channel thresholds are computed and set to the pedestal mean + 3$\sigma$.
On the left of figure~\ref{PlotCalib} a typical distribution of the electronic noise is shown for the 1600 channels in the CHERCAM plane. The noise level, given by the pedestal width, is found to be at an average of 5 ADC counts in gain x5.

In a second calibration run, lasting about 50\,s (at a 200\,Hz rate), the LED pulser is used. With the LED tuned to produce in average of the order of 1 photoelectron per channel and event, single photoelectron spectra are taken. This calibration run is used to monitor an eventual gain evolution during the flight and also to check the health of the PMTs.
From run sizes of 10\,k events the gain of each channel can be estimated by removing the pedestal contribution and simply computing the mean value of each spectrum.

The map of a typical gain distribution across the CHERCAM plane is shown on the right of figure~\ref{PlotCalib}.
The typical single photoelectron gain is 25 ADC channels. Six dead channels are visible as well as a few channels where the gain is higher than expected, which is due to PMT gain sorting errors.

During the detector construction phase, the pedestal mode was also used for
precisely locating possible residual light leaks.

\section{Radiative plane}
\label{AerogelSection}
\begin{figure}[th]
\begin{center}
\includegraphics[width=10cm]{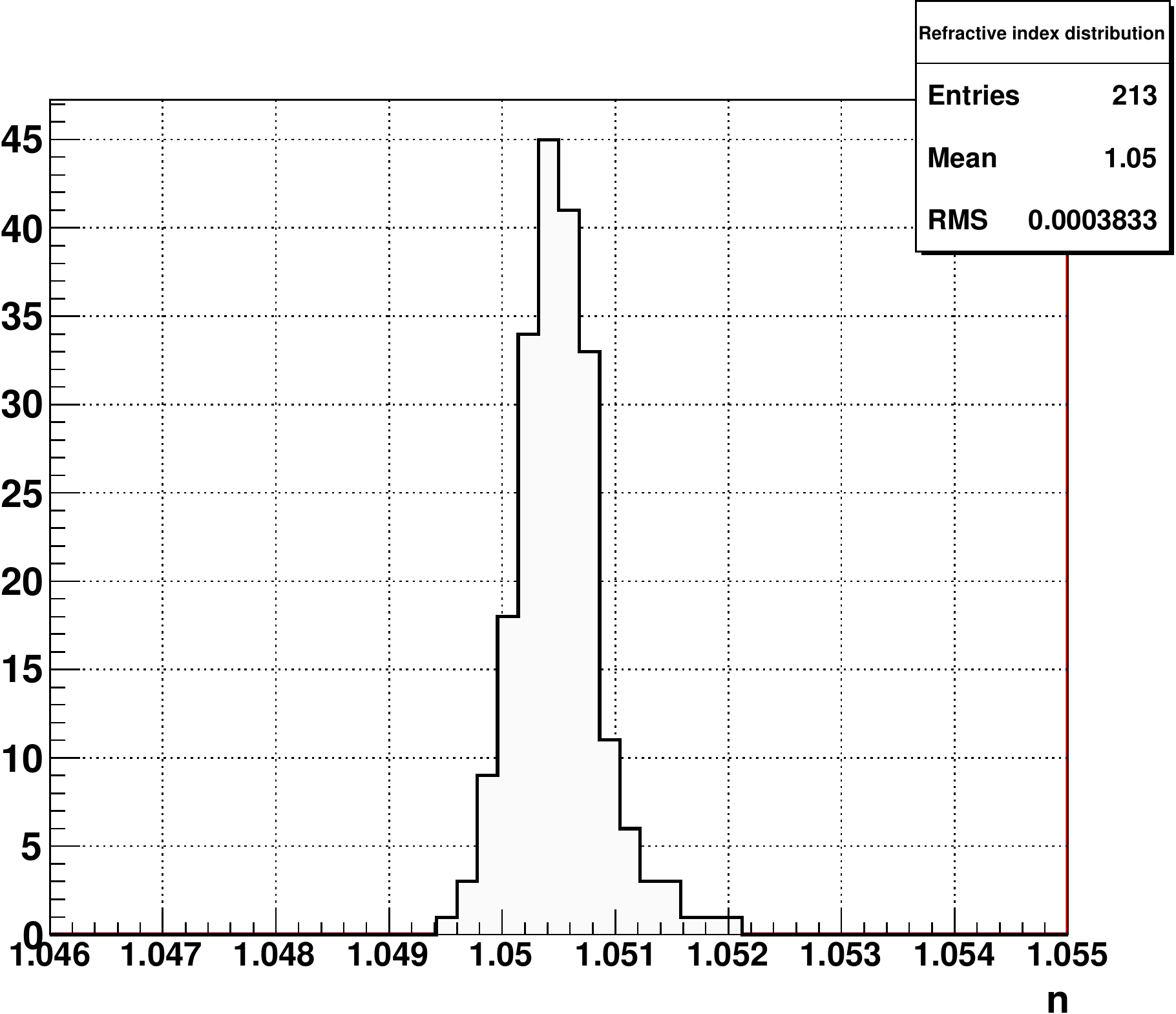}
\caption{Histogram of the refractive showing the good uniformity of the index distribution over the whole CHERCAM aerogel plane.}
\label{SP50index_fit}
\end{center}
\end{figure}

The Cherenkov radiator consists of a 20.8\,mm thick silica aerogel plane. Each cell consists of a stack of two Matsushita-Panasonic SP50 aerogel tiles of size $\rm 105 \ mm\ \times 105 \ mm  \times 10.4 \ mm$, with a refraction index of n=1.05. Note that the tile size matches the dimensions of a submodule and that they are hygrophobic.
The characterization of the tiles was recently reported in \cite{AGL10}. The refraction index uniformity is shown in figure~\ref{SP50index_fit}. The average refractive index is 1.0505 with a RMS of the distribution of only 0.0004. 
Additionally, the tiles clarity coefficient has been measured on four samples at the Madrid CIEMAT laboratory. The measured values were all within the range $\rm 15-17 \times 10^3$ \textmu m\textsuperscript{4}/cm.
 
\begin{figure}[th]
\begin{center}
\includegraphics[width=10cm]{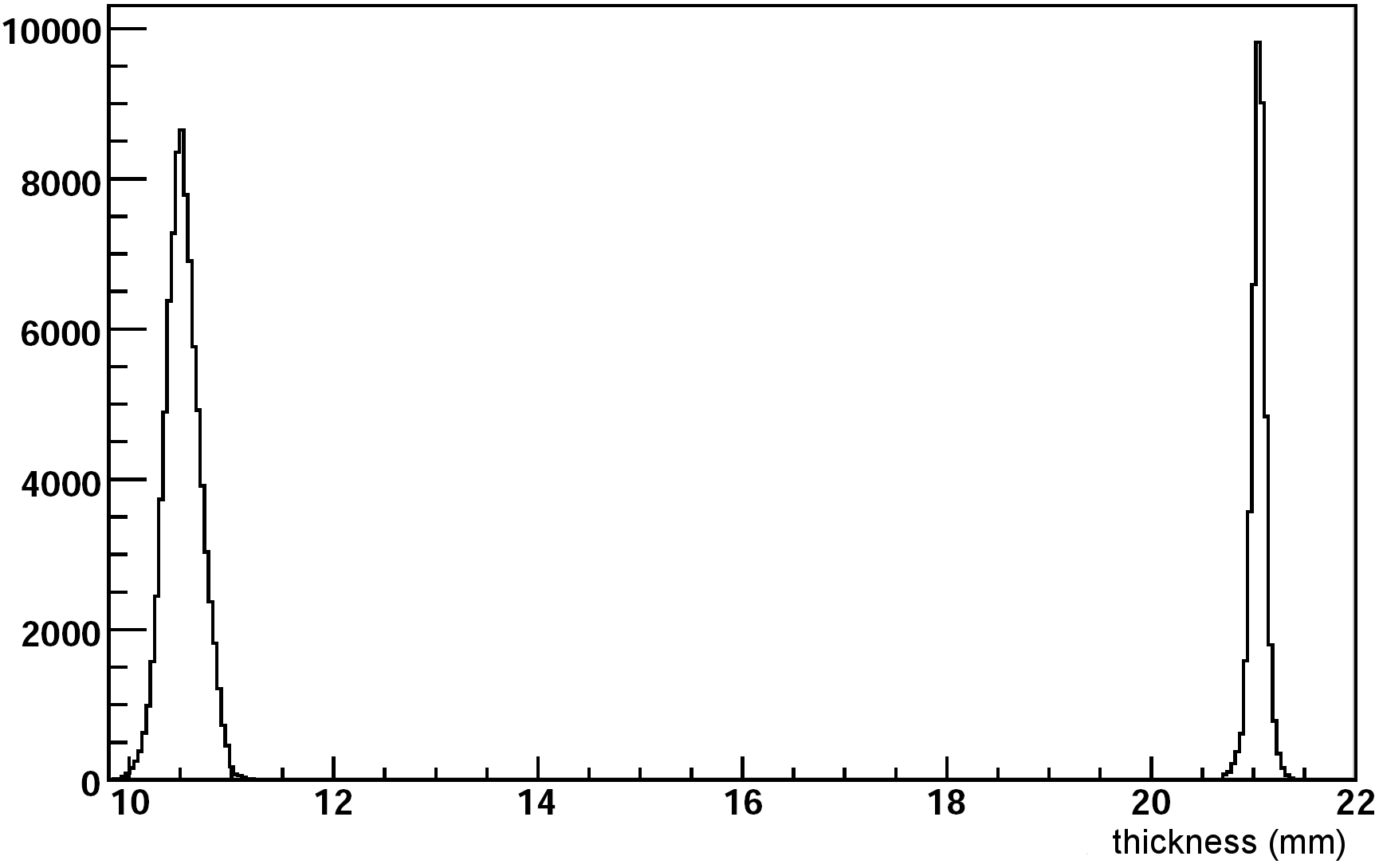}
\caption{Histogram of individual tiles and stacked tiles thickness distribution.}
\label{AppariemmentAGL}
\end{center}
\end{figure}

The tiles have been sorted by thickness and pairs have been matched to equalize the thickness between the aerogel stacks. This way the variation of the Cherenkov photon yield across the radiator plane is minimized.
Figure \ref{AppariemmentAGL} shows the thickness distribution of the individual tiles (left peak) and of the matched pairs on the same histogram (right peak). 
While the tiles are produced with a standard deviation of sigma tile = 172\,\textmu m for the tile thickness, in the formed stacks the standard deviation could be reduced by 60\% to sigma stack = 68\,\textmu m \cite{theseYO}.

\begin{figure}[th]
\begin{center}
\includegraphics[width=15cm]{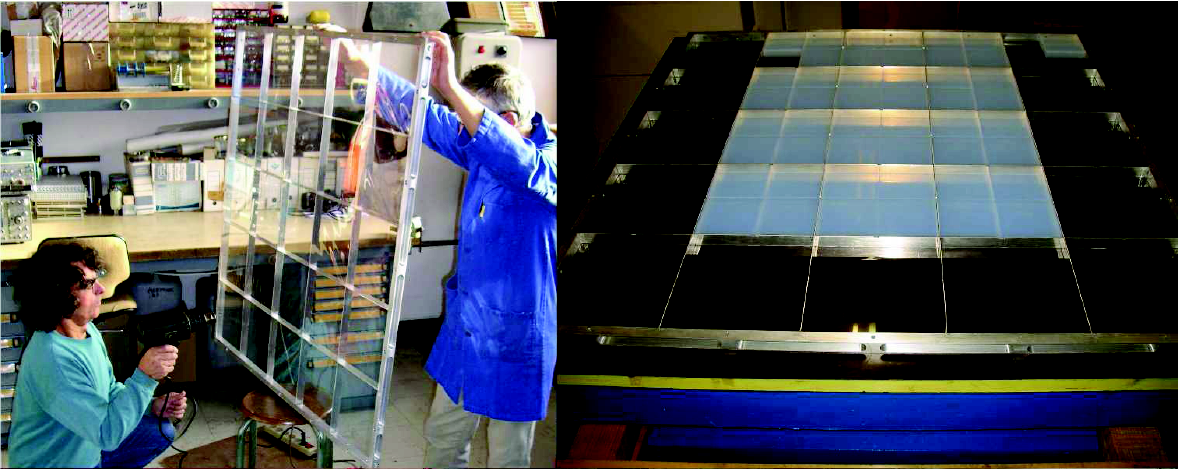}
\caption{Picture of two steps of the aerogel installation procedure. Left picture shows the 50\,\textmu m thick thermostretched foil of polyester frame being glued along the edges of the frame aerogel. Right picture shows 12 sub-arrays of $2 \times 2$ stacks of aerogel installed in the aerogel frame. }
\label{InstallAGL}
\end{center}
\end{figure}

The stacks of aerogel tiles were mounted in the Aerogel frame in a layout which matches that of the array of photon detector modules below. Figure~\ref{InstallAGL} shows two steps of the installation procedure. At first, a single 50\,\textmu m thick thermostretched foil of polyester \footnote{PMX726 from HIFI} is glued along the edges of the frame which will be facing the PMT plane. The foil is used to ensure the enclosing of the tiles in their housing. The material used to this purpose has a transmission coefficient ensuring $\sim$80-85\% light transmission in the wavelength range 320\,nm to 800\,nm \cite{theseYO}. In the next step, $2 \times 2$ stacks of aerogel were directly installed in each frame lodging, each cell size matching perfectly the size of the four grouped stacks. 
Each stack was backed with a black sheet of thin cardboard and with a foil of thin foam that fills the mechanical void and gently locks the tiles in their confinement volume. The aerogel frame was closed on the top side by the upper honeycomb lid of the detector, where the cosmic particles enter. The lid of detector was finally installed on CHERCAM. The black cardboard absorbs photons from Rayleigh scattering in the aerogel and Cherenkov light which was reflected at the polyester foil, thereby reducing the background from photons with no or poor relation to the charge of the primary particle. Note that a very limited volume is captured in the aerogel frame and that it is not gas tight, thus there is no structural issue due to pressure equalization during the balloon ascension phase.


%
\section{Vacuum and thermal tests}
\begin{figure}[th]
\begin{center}
\includegraphics[angle=-90,width=10cm]{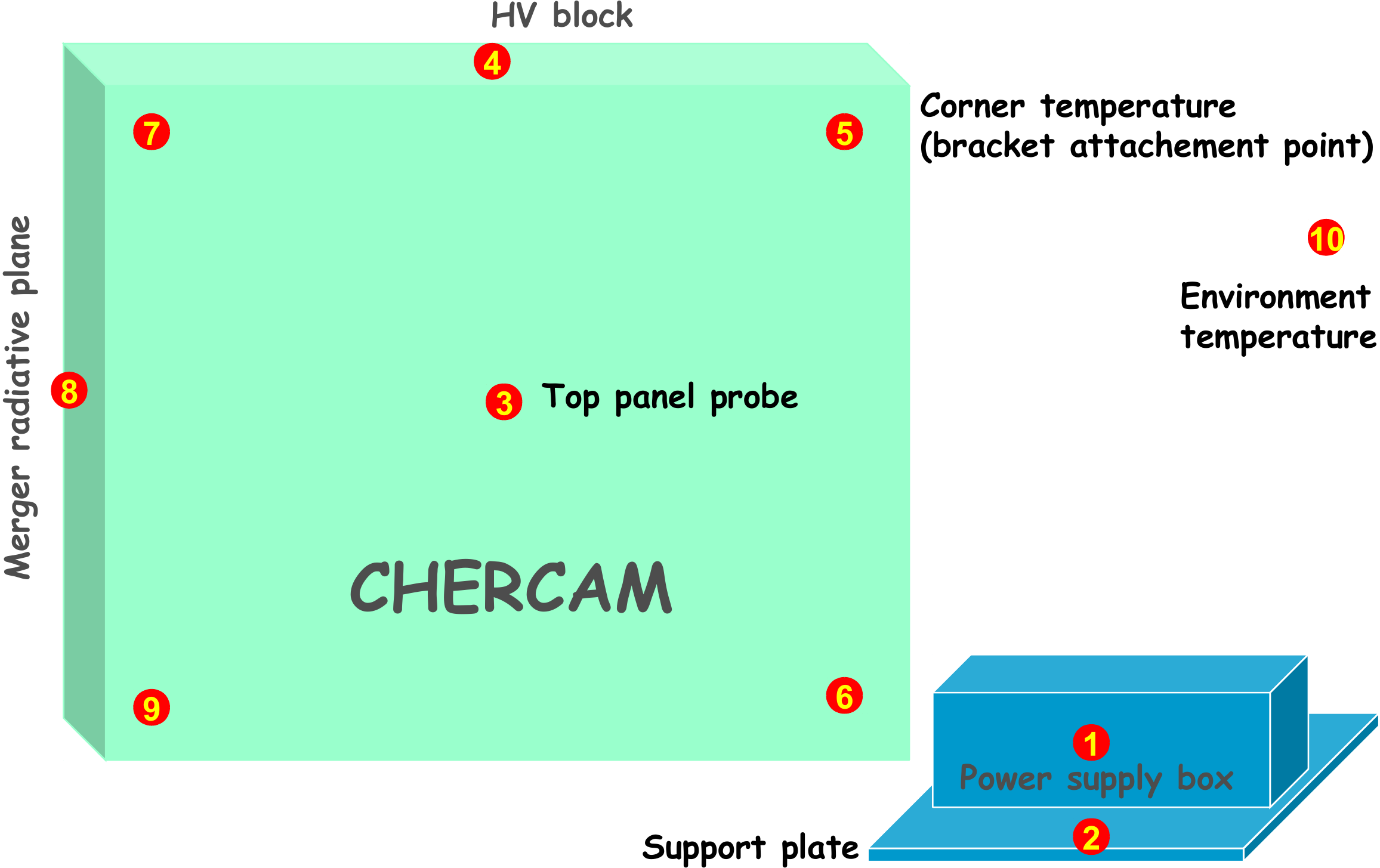} \\
\vspace{0.5cm}
\includegraphics[angle=-90,width=14cm]{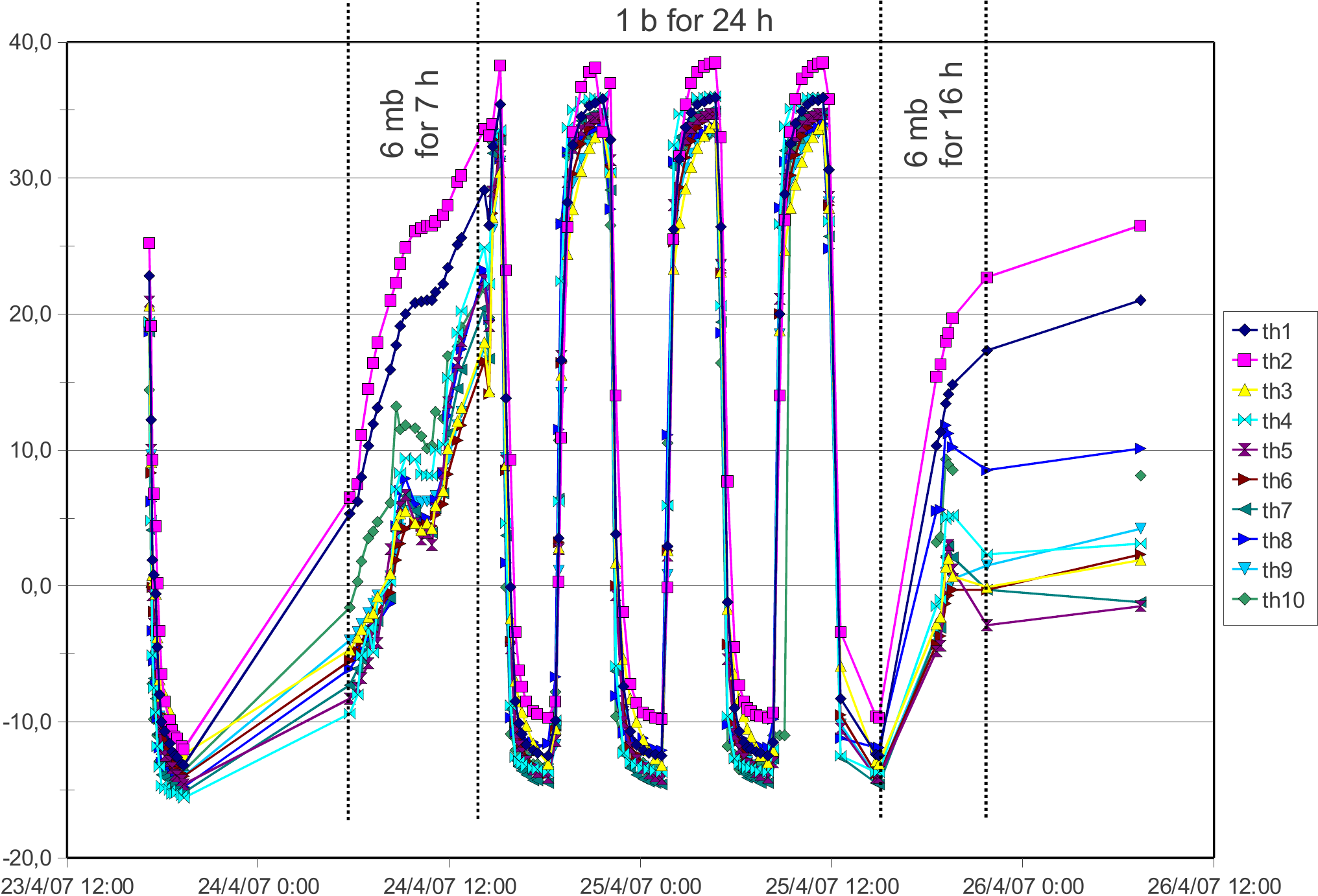}
\caption{Top: external temperature probe locations. Bottom: Temperature variation at probe locations as a function of time (day/hour), the ambient pressure of the 3 phases of the test are marked on the plot.}
\label{IntespaceTemp}
\end{center}
\end{figure}
At the early stage of the development, during the R\&D phase, extensive testing of the potting procedure of the submodules and high voltage converters in low pressure vacuum chambers were conducted. The results lead to design modifications of the HV distribution board and to a new injection procedure. For instance, it appeared that during potting material injection at the base of the PMTs, a large air bubble systematically formed on the opposite side leading to voltage breakdown. A modified PCB version with a hole at the opposite side of injection solved the problem, the hole location being chosen in order to be close to unconnected PMTs pins.

With CHERCAM fully build up the detector was commissioned in two series of tests, each recreating aspects of the experimental conditions during balloon flight.

Using the vacuum test facility of the Institut d'Astrophysique Spatiale (IAS) \cite{IAS} at Orsay CHERCAM was operated at vacuum conditions of 5\,mbar, but kept at room temperature which is equivalent to ambient temperature inside the payload during the flight. Cosmic ray as well as calibration runs were taken under these conditions for 48 hours, thereby also validating the thermal model of the system. No breakdown of the HV occurred over that period.

The second test was conducted at the INTESPACE Toulouse facility \cite{INTESPACE} in a vacuum and temperature vessel. After verifying CHERCAM proper operation at 6\,mb for 7\,hours, it was taken back to atmosphere pressure and subjected to four thermal cycles between -15\textcelsius \  and +35\textcelsius. After this cycling, it was verified for 16\,hours at 6\,mb that no HV breakdown occurred. With this test it was verified that the insulation properties of the potting around the PMTs, on the FE boards and around the HVPS boards do not degrade due to the thermal expansion when the system is subjected to temperature changes.  Cosmic ray as well as calibration runs were taken regularly during the whole testing.
The position of the ten external temperature probes used during this test are shown in the upper part of figure~\ref{IntespaceTemp}. The lower part of the figure documents the four temperature cycles which the system was subjected to over the course of $\sim$48\,hours.

%

\section{Summary and conclusion}
The Cherenkov imager CHERCAM was designed and built for the charge measurement of nuclear cosmic-ray particles in the CREAM instrument aboard stratospheric balloon flights. 
This was the first application of the Cherenkov imaging technique for charge measurement in embarked experiments.

The CHERCAM design was tightly constrained by the requirements of balloon-borne experiments, namely power and weight limitations, and thermal aspects. The main difficulty lied in the high voltage handling in the detector at low atmospheric pressure conditions. Therefore, the design was based on technical solutions proven to be suitable for extreme condition operations.

The construction phase including the validation tests lasted about one year. The first successful flight of the CHERCAM detector was accomplished with the CREAM-III experiment in 2007.
So far, the detector has been operated successfully during four consecutive yearly flights over Antarctica, during which about 100 effective days of measurement were accumulated. After each payload recovery, the re-integration of CHERCAM into the CREAM experiment for the following flight was possible thanks to very short refurbishment phases.

The mechanical and electronic layout of CHERCAM has proven to be well suited for balloon-borne experiments: no PMTs failed, no HV breakdown events were recorded, neither any single event latch-up nor any single event upset occurred, data acquisition and slow control went reliably and smoothly and finally, no degradation of the properties of the aerogel tiles was observed nor were they damaged by the mechanical forces, only a few non-significant cracks appeared on a small set of tiles.

Preliminary results of the detector performances were presented in \cite{derome} and demonstrated that the conceptual design is relevant for cosmic-ray physics. The in-flight performances of the detector will be presented in a forthcoming paper. 

\section*{Acknowledgments}
The CHERCAM design and construction was supported by IN2P3/CNRS and the ANR-06-BLAN-0042 funding. 
This work was supported in the U.S. by NASA grant NNX08AC11G and its predecessor grants. The authors wish to thank the Institut
d'Astrophysique Spatiale (IAS) Orsay for providing access to their vacuum test
facility. 


\end{document}